\documentclass[11pt]{article}

\usepackage{jcappub} 

\usepackage[T1]{fontenc} 

\usepackage{amssymb}
\usepackage{graphicx, epsfig, bm, amsmath,mathtools}

\usepackage{hyperref}
\usepackage{cleveref}

\usepackage{subcaption}
\usepackage{soul}
\usepackage{float}
\usepackage{color}

\newcommand{\bea}{\begin{eqnarray}}
\newcommand{\eea}{\end{eqnarray}}
\def\[{\left[}
\def\]{\right]}
\def\({\left(}
\def\){\right)}

\def\d{{\rm d}}
\def\Mpl{M_{\rm Pl}}

\title{Twisting inflation to sub-Planckian axion decay constants}
\author[a,b]{Peter Adshead,}
\author[c,d]{Suddhasattwa Brahma,}
\author[a]{Indranil Das}

\affiliation[a]{Illinois Center for Advanced Studies of the Universe \& Department of Physics, University of Illinois at Urbana-Champaign, Urbana, IL 61801, U.S.A.}
\affiliation[b]{Department of Physics and Astronomy,
University of Pennsylvania, 209 South 33rd St, Philadelphia, PA 19104}

\affiliation[c]{Higgs Centre for Theoretical Physics, School of Physics and Astronomy,
University of Edinburgh, Edinburgh EH9 3FD, UK}
\affiliation[d]{Physics and Applied Mathematics Unit, Indian Statistical Institute, 203 B.T. Road, Kolkata 700108, India}

\emailAdd{adshead@illinois.edu}
\emailAdd{suddhasattwa.brahma@gmail.com}
\emailAdd{idas3@illinois.edu}

\abstract{We study pseudoscalar inflation in the Einstein-Cartan-Palatini (first-order) formulation of gravity while allowing for torsion. We introduce two non-minimal interactions in the gravitational sector --- pseudoscalar couplings to the Pontryagin density (Chern-Simons term) and the Nieh–Yan topological invariant. In the presence of these terms the rolling pseudoscalar sources non-trivial torsional fields during inflation. We show that pathological gradient and ghost instabilities limit the strength of the coupling to the Pontryagin density during inflation. Furthermore, we show that the interaction with the Nieh-Yan term induces a new contribution to the pseudoscalar kinetic term which parametrically increases its decay constant and allows for inflation on steep potentials. The torsion field generated by the background is parity violating, which is manifest in the resulting chiral gravitational wave spectrum. We find that the scalar sector is largely unaffected beyond the remapping of the axion decay constant to a larger value. Consequently, we demonstrate that Generalized Natural Inflation, D-brane models, and Hilltop Squared Inflation can satisfy current observational constraints with sub-Planckian decay constants. 
}

\begin{document}
\maketitle
\flushbottom

\section{Introduction}\label{section introduction}

Inflation \cite{Guth:1980zm, Linde:1981mu, Albrecht:1982wi, Linde:1983gd} is now well-established \cite{Planck:2018jri} as the leading theory to fix the initial state of the primordial universe. While the current observations --- red-tilted, Gaussian scalar fluctuations, with negligible gravitational waves ---  are well-accounted for by inflation driven by a single scalar field rolling slowly on a concave potential, the fundamental theory of inflation is yet unknown. 

Most existing inflationary scenarios drive inflation with a scalar field evolving on a sufficiently flat potential. Much of contemporary model-building involves crafting mechanisms for the origin of such a flat  (effective) potential that are protected from uncontrolled quantum corrections that lead to a large inflaton mass (the so-called $\eta$-problem). A popular way of avoiding the large quantum corrections is to utilize the shift-symmetry of a pseudoscalar \cite{PhysRevLett.65.3233, Kinney:1995xv, Arkani-Hamed:2003wrq, Dimopoulos:2005ac, Easther:2005zr, Silverstein:2008sg, McAllister:2008hb, Berg:2009tg, Adshead:2012kp, Adshead:2012qe,Martinec:2012bv, Long:2014dta, McAllister:2014mpa, Adshead:2016omu}. Pseudoscalar, or axion-like particles, either from some unified gauge group or from string theory, possess shift symmetries and are therefore ideal candidates for the inflaton. Non-perturbative corrections in the form of instantons lead to the breaking of the continuous shift symmetry to a discrete one $\vartheta \rightarrow \vartheta + 2\pi f$, where the `axion decay-constant' $f$ plays a crucial role in the phenomenology of these models. It is now well-established, either due to the `no global symmetries in quantum gravity' conjecture for gauge theory axions or by the `Weak Gravity Conjecture' for controlled string theory constructions, that super-Planckian decay constants cannot be obtained within a well-controlled UV-complete EFT \cite{Banks:2003sx, Heidenreich:2015wga}. This presents a dilemma for many pseudoscalar-driven models, as this is often precisely what is needed for making these models compatible with current observations of the cosmic microwave background (CMB). Significant model building has gone into sidestepping this issue using, for example, monodromies \cite{Silverstein:2008sg, McAllister:2008hb, Berg:2009tg}, large numbers of fields \cite{Dimopoulos:2005ac}, or aligned directions in multifield scenarios \cite{Long:2014dta, McAllister:2014mpa}, to extend the field range over multiple cycles of the underlying axion(s).

In this paper, we study gravitational terms that couple naturally to pseudoscalars --- the Pontryagin (or Chern-Simons) and Nieh-Yan topological densities --- and demonstrate that they can potentially alleviate this problem, and allow for sub-Planckian axion decay constants that support viable models of inflation. Inflation generically results in a near-isotropic and homogeneous Friedmann-Lema\^itre-Robertson-Walker (FLRW) background spacetime on which these terms vanish. We therefore require additional degrees of freedom in the gravitational sector for these interactions to play a role in the dynamics of the classical background. We show that these additional degrees of freedom can be found by allowing for torsion in the system.

Given this context, we work in the first-order, Einstein-Cartan-Palatini formulation of general relativity, where the tetrad and the spin-connection are the fundamental variables. In the absence of torsion, the theory reduces to the standard metric version of Einstein's General Relativity. We extend the Einstein-Cartan gravitational action with Pontryagin (Chern-Simons) and Nieh-Yan densities that allow a natural, albeit non-minimal, coupling to the pseudoscalar. This is a fundamental point of difference from non-minimally coupled versions of Higgs inflation that have been previously studied in the presence of similar terms --- we only keep shift-symmetric operators. Specifically, we study pseudoscalar inflation in the presence of the Pontryagin and the Nieh-Yan topological densities, while allowing for torsion when rendering them dynamical. 

On the one hand, we show that the Chern-Simons term, which is a dimension-$5$ operator after coupling to a canonically normalized pseudoscalar, must have an extremely large coupling in order to play a role in the background dynamics. The validity of an effective field theory with such a large coupling notwithstanding, we demonstrate that ghost and gradient instabilities make the theory inviable. On the other hand, the dimension-$2$ Nieh-Yan density \cite{Nieh:1981ww}, (which leads to a dimension 3 operator when coupled to the canonically normalized axion field) leads to an effective theory for the background pseudoscalar with kinetic term, after integrating out the torsion fields,  $(1+6 M^4/\Mpl^2f^2)(\partial\vartheta)^2$, where $M$ is the mass scale associated with the Nieh-Yan density, $f$ is the axion decay constant, and $\Mpl$ is the Planck mass.

Torsion has been studied already in a number of cosmological contexts (see, for example, \cite{Poplawski:2010kb, Lu:2016bcx, Akhshabi:2017lyg, Cid:2017wtf, Kranas:2018jdc, Kasem:2020ddi, Guimaraes:2020drj, Cite1, Cite2, Cite3, Cite4, Cite5, Shaposhnikov_2021, Shaposhnikov_2020, he2024quantumcorrectionshiggsinflation, Pradisi_2022, Di_Marco_2024, He_2024, Salvio_2022, he2025increasensregularizedpole, L_Buchbinder_1990, Shapiro_2014, Shapiro_2002, he2025torsioninducedcurrentscalaroncoupling}). The work of ref.\ \cite{Poplawski:2010kb, Kranas:2018jdc} formulates cosmology in the Einstein–Cartan–Kibble–Sciama theory of gravity, which extends general relativity to the spin of matter.  The work of ref.\ \cite{Cid:2017wtf} studies non-minimally coupled scalar cosmologies and relaxes the null torsion constraint. The work of ref.\ \cite{Lu:2016bcx} introduces a similar ansatz for the torsion tensor, however, they work within the context of Poincare gauge gravity, while the authors of ref.\ \cite{Kasem:2020ddi} source torsion via a gauge field texture. Furthermore,  refs.\ \cite{Cite1, Cite2, Cite3, Cite4, Cite5} build pseudoscalar inflationary models with  a dynamical affine connection. Our work differs from these and others; we study inflationary spacetimes in Einstein-Cartan gravity sourced by a rolling pseudoscalar on a suitable potential. Torsion in our model is sourced by the interaction of the pseudoscalar with the topological Nieh-Yan and Pontryagin densities. 

This paper is organized as follows. In section \ref{section gravity in first order form}, we briefly review the Einstein-Cartan-Palatini formulation of general relativity, introduce our model and derive the general equations of motion for the system. In section \ref{section cosmological solutions}, we work with a Friedmann-Robertson-Walker background and study the resulting equations of motion. In section \ref{section perturbations}, we study scalar and tensor perturbations about the classical background solutions. We explore the phenomenology of our model in section \ref{results}, before concluding in section \ref{section conclusions}. Details of our calculations are relegated to appendices. In appendix \ref{Appendix reduction of the general torsion tensor to the effective one}, we prove that our ansatz for the torsion tensor is self-consistent, while appendices \ref{Appendix Expression for Tensor power spectrum} and \ref{Appendix Equation of motion for the scalar perturbation} contain the details of the computations of the spectra of scalar and tensor fluctuations.
 
We work in natural units where $\hbar = c = 1$, and we retain the Planck mass $M_{\rm Pl}^2 = (8\pi G_{\rm N})^{-1}$. Internal Lorentz indices are denoted by capital Latin letters ($A, B, \ldots$), and the spacetime indices, denoted by the Greek letters ($\mu, \nu$, \ldots), both run from $0$ to $3$. The Levi-Civita symbol is denoted  $\epsilon_{ABCD}$, with $\epsilon_{0123} = 1$, and we work throughout in the $(+---)$ signature.
 
%
\section{Gravity in Einstein-Cartan-Palatini form}\label{section gravity in first order form}
%

We work in the Einstein-Cartan-Palatini (first-order) formulation of gravity to allow for torsion. We define a local reference frame at each point of the $(3+1)$-dimensional manifold $\mathcal{M}$, the tetrad $e^A_\mu$, such that the metric can be written as $g_{\mu\nu}=e^A_\mu e^B_\nu \eta_{AB}$, with $\eta_{AB}$ being the flat Minkowski metric on the internal space of coordinates. The metrics $g_{\mu\nu}$ and $\eta_{AB}$ can raise or lower the spacetime and internal tangent-space indices, respectively. 

The gravitational action can be reformulated in the first-order form as a function of the tetrad $\(e^A\)$  and spin-connection variables $\(\omega^{AB}\)$. Both of these are 1-forms on the manifold $\mathcal{M}$. Instead of using the abstract index notation, in what follows we suppress the spacetime indices while retaining wedge products between forms to simplify the notation.

We start with the gravitational action in the form  (see, for example, \cite{DAuria:1981ddz, Leigh:2007wf, Leigh:2008tt})
\bea
\mathcal{S}_{\rm EC}=-\frac{\Mpl^2}{4}\int \epsilon_{ABCD} \, e^A\wedge e^B\wedge R^{CD}\,,\label{EH action equation}
\eea
where the curvature 2-form is
\begin{align}\label{curvature 2 form equation}
R^{AB} =\d\omega^{AB} + \omega^A{}_C\wedge\omega^{CB}\,.
\end{align}
Since we envisage an axion-driven model of inflation, we add a minimally-coupled pseudo-scalar, $\vartheta$, with an as-yet unspecified  potential, given by the action
\begin{align}\label{scalar action equation}
\mathcal{S}_{\vartheta} = \int\( \frac{1}{2}\star\! \d \vartheta \wedge \d\vartheta -\frac{\epsilon_{ABCD} e^A \wedge e^B\wedge e^C\wedge e^D}{3!}\, V(\vartheta)\)\,.
\end{align}
In order to generate magnetic-type, or derivative interactions with the pseudo-scalar field, we couple the axion to the gravitational field via a Chern-Simons-type coupling \cite{Jackiw:2003pm, Alexander:2009tp}
\bea\label{CS action}
\mathcal{S}_{\rm CS}=\frac{\tilde{\alpha}}{4 f}\int\d\vartheta\wedge \left( \omega^{AB} \wedge\d\omega_{AB} +\frac{2}{3}\omega^A{}_B\wedge\omega^B{}_C\wedge\omega^C{}_A\right)\,,
\eea
where we have dropped a boundary term and $\tilde{\alpha}$ denotes the dimensionless coupling parameter. Similarly, we consider the effect of coupling the pseudoscalar to the Nieh-Yan form \cite{Nieh:1981ww}
\begin{align}\label{NY action}
S_{\rm NY} = -nf\int \d\vartheta \wedge T^A \wedge e_A\,,
\end{align}
where $n$ is a dimensionless coupling and $f$ is the axion decay constant. We have introduced the torsion two-form $T^A$ as
\begin{align}\label{eqn torsion definition}
T^A =\d e^A + \omega^A{}_B\wedge e^B\,.
\end{align}
In eq.\ \eqref{NY action}, $n = M^2/f^2$ measures the ratio of the new mass scale to that of the axion decay constant,  where $M$ is a new ultraviolet (UV) mass scale associated with the Nieh-Yan density.  Since we anticipate the Nieh-Yan density to have a gravitational origin, we expect $M$ to potentially be of the same order of magnitude as $\Mpl$. However, we remain agnostic about this and let $n$ vary to see which regions of its parameter space allow for inflationary solutions. As it turns out, this is a key reason why we are able to find inflationary solutions with sub-Planckian decay constants, as this implies $n>1$ is an $\mathcal{O}(10-100)$ number.

%
\subsection{Equations of motion} \label{section equations of motion}
%

Before looking for cosmological solutions, we first compute the general equations of motion for the system. To obtain these, we vary the action with respect to the independent degrees of freedom; here, these are the axion, tetrads, and spin-connection fields.

Varying with respect to the tetrad yields the Einstein equations 
\bea\label{einstein equation CS and NY}
-\frac{\Mpl^{2}}{2}R^{AB} \wedge e^{C} \epsilon_{ABCD} - 2nf\d\vartheta \wedge \d e_{D} - 2nf  \d\vartheta \wedge \omega_{DC} \wedge e^{C} = \tau_D\,,
\eea
where $\tau_D$ is the stress-energy 3-form, which is related to the stress-energy tensor via 
\bea\label{stress energy 3 form and tensor equation}
\tau_{D} = \frac{1}{3!} \Big[ \partial_{D}\vartheta \partial^{\beta}\vartheta - e^{\beta}_{D} \Big( \frac{1}{2} \partial_{\mu}\vartheta \partial^{\mu}\vartheta - V(\vartheta) \Big) \Big] \epsilon_{\beta ABC} e^{A} \wedge e^{B} \wedge e^{C}\,. 
\eea 
Varying with respect to the spin-connection yields the torsion constraint
\bea\label{torsion constraint CS and NY equation}
\epsilon_{ABCD}T^C \wedge e^D = \frac{\alpha}{\Mpl^2}\d\vartheta \wedge R_{AB} - \frac{2nf}{\Mpl^{2}} \d\vartheta \wedge e_{A} \wedge e_{B}\,,
\eea
where the torsion tensor is
\begin{align}
T^A = \d e^A + \omega^A{}_B\wedge e^B ,
\end{align}
and we have redefined $\alpha = \tilde{\alpha}/f$. 

Finally, varying with respect to the scalar yields the usual Klein-Gordon equation, with new source terms. In a coordinate basis, this reads
\begin{align}\label{KG equation CS and NY}\nonumber
\partial_{\mu}\(\sqrt{-g}g^{\mu\nu}\partial_{\nu}\vartheta\) + \sqrt{-g}\frac{d V}{d\vartheta} = \frac{\alpha}{4} \epsilon^{\mu\nu\alpha\beta} \partial_{\mu}\[ \hspace{2pt} \omega_{\nu A}\hspace{0.5pt}^{B} \( \frac{1}{2}R_{\alpha\beta B}\hspace{0.5pt}^{A} - \frac{1}{3}\omega_{\alpha B}\hspace{0.5pt}^{C} \omega_{\beta C}\hspace{0.5pt}^{A} \)\]
\\
+ \frac{nf}{2}\epsilon^{\mu\nu\rho\sigma} \partial_{\mu}(T^{A} {}_{\nu\rho} e_{A\sigma}) \,.
\end{align}
Notice here that in the absence of the Chern-Simons or Nieh-Yan coupling (or other sources of torsion, such as fermions),  the torsion two-form vanishes, giving the relation between the tetrads and the spin connection
\begin{align}\label{torsionless equation}
T^A = \d e^A + \omega^A{}_B\wedge e^B = 0\,,
\end{align}
leading to a metric compatible connection. 
%
\section{Cosmological solutions}\label{section cosmological solutions}
%
To be consistent with cosmological observations, the background must describe a homogeneous and isotropic universe. We therefore choose the following ansatz for the tetrads
\begin{align}\label{vielbeins}
e^0  = \d t, \quad e^b = a(t)\d x^b\,,
\end{align}
which leads to the FLRW metric
\begin{align}\label{FRW unperturbed metric}
\bar{g}_{\mu\nu}\d x^\mu \d x^\nu= \d t^2 - a^2(t)\d {\bf x}^2\,, 
\end{align}
where $a(t)$ is the usual scale factor for the universe.\footnote{From now on, we split up the internal indices $B$ as $(0,b)$, where the small Latin alphabets label the spatial Lorentz indices while $0$ stands for the temporal one.} On the FLRW background, the inflaton field is assumed to be homogeneous, $\vartheta = \vartheta(t)$, as usual.

We introduce the most general ansatz for the torsion 2-form 
\bea\label{general ansatz torsion}
T^0 &=& 0\,,
\\
T^i &= & h(t)e^0\wedge e^i + h^i_{j}(t)e^0 \wedge e^j - \phi(t)\epsilon^i_{jk} e^j \wedge e^k+f^i{}_{l}\epsilon^l_{jk} e^j \wedge e^k\,,
\eea
where $h^{ij}$ and $f^{il}$ are traceless. After imposing the equations of motion, this reduces to (see appendix \ref{Appendix reduction of the general torsion tensor to the effective one} for details)
\bea \label{ansatz torsion}
T^0 &=& 0\,,
\\
T^i &= & h(t)e^0\wedge e^i - \phi(t)\epsilon^i_{jk} e^j \wedge e^k.
\eea
The spin-connection can be split into a torsion-less and torsion-full parts
\bea\label{spin connection two parts}
\omega^{AB} = \bar{\omega}^{AB}+\tilde\omega^{AB}\,.
\eea
The torsion-less spin connection ($\bar{\omega}^{AB}$) generates the usual metric compatible Christoffel connection with components
\begin{equation}\label{eqn:spinnotorsion}
\bar{\omega}^{a}\hspace{0.5pt}_{0} =  H(t)\, e^a,\quad
\bar{\omega}^{a}\hspace{0.5pt}_{b}  =  0\,,
\end{equation}
where $H(t)=\dot{a}/a$ is the Hubble parameter. The torsion-full parts of the spin-connection, $\tilde\omega^{AB}$, are
\begin{align}\label{eqn:spintorsion}
\tilde\omega^{0i} =& h(t)e^i \,,\quad 
\tilde\omega^{ij} =   \phi(t)\epsilon^{ij}{}_{k}e^k \,.
\end{align}
With the tetrads in eq.\ \eqref{vielbeins}, and the spin connection given by eqs.\ \eqref{eqn:spinnotorsion} and \eqref{eqn:spintorsion}, the curvature two-form has components
\bea\label{curvature 2 form components}
R^{0i} &=& \[(\dot{H}+\dot{h})+H(H+h)\]e^{0}\wedge e^i+\phi(H+h)\epsilon^i{}_{jk}e^{j}\wedge e^{k}\,,\\
R^{ij} &=& \[(H+h)^2-\phi^2\]e^i \wedge e^j + (\dot\phi + H\phi)\epsilon^{ij}{}_{k}e^k \wedge e^0\,.
\eea
where we have used the definition of the full curvature tensor from eq.\ \eqref{curvature 2 form equation}.

\subsection{Pseudoscalar inflation in dynamical Chern-Simons gravity}\label{section eom chern simons}

We start by considering the effect of the Pontryagin density on the inflaton field in the presence of torsion. Substituting the above ansatz for the spin-connection into the equations of motion in eq.\  \eqref{einstein equation CS and NY} (setting $n = 0$), and assuming that the background $\vartheta$ depends only on time, we obtain
\begin{align}\label{eom CS}
(H+h)^{2} - \phi^{2} &= \frac{1}{3\Mpl^{2}}\(\frac{\dot{\vartheta}^{2}}{2} + V(\vartheta)\),
\\
2(\dot{H} + \dot{h} + H(H+h)) + (H+h)^{2} - \phi^{2} &= -\frac{1}{\Mpl^{2}}\( \frac{\dot{\vartheta}^{2}}{2} - V (\vartheta)\),
\\\label{eqn:torsion equation}
\phi &=  \frac{\alpha\dot{\vartheta}}{2\Mpl^{2}}[(H+h)^{2} - \phi^{2}],
\\
h &= -\frac{\alpha \dot{\vartheta}\phi(H+h)}{\Mpl^{2}}\,,
\end{align}
while the sourced Klein-Gordon equation, eq.\  \eqref{KG equation CS and NY}, for the pseudoscalar (setting $n = 0$) takes the form
\begin{equation}\label{KG equation CS}
\ddot{\vartheta} + 3H\dot{\vartheta} + \frac{dV}{d\vartheta} =-  3\alpha[\dot{\phi}\big((H+h)^{2} - \phi^{2} \big) + 2\phi(H+h)(\dot{H} + \dot{h}) + 3H\phi(H+h)^{2} - H\phi^{3} ]\,.
\end{equation}
Notice that in the presence of torsion, the Klein-Gordon equation \eqref{KG equation CS} has a non-trivial, but classical, source term on the right hand side. When $\alpha = 0$, $\phi = h = 0$, and the system reduces to the standard single field inflation scenario in Einstein gravity.

We can estimate the magnitude of the dimensionless coupling $\alpha$ in order for the Pontryagin density to play an important role in the dynamics of the background. Consider the first slow-roll parameter,
\begin{align}
\epsilon_{H}= -\frac{\dot{H}}{H^2},
\end{align}
which, in the standard slow-roll approximation ($\dot{\vartheta}^2 \ll V$ and $\ddot{\vartheta} \ll V'$), can be written 
\begin{align}\label{epsilon H chern simons}
\epsilon_{H} \simeq -\frac{36 \dot{\vartheta} \Mpl^{10}}{\left(V \left(12 \Mpl^6+\alpha ^2 V \dot{\vartheta}^2\right)\right)^{3/2}}\[\frac{3 \alpha ^2 V^2\ddot{\vartheta} \left(10 \Mpl^6+\alpha ^2 V \dot{\vartheta}^2\right)}{\left(6 \Mpl^6+\alpha ^2 V \dot{\vartheta}^2\right)^2} + 2V' - \frac{36V'\Mpl^{12}}{(6 \Mpl^6+\alpha ^2 V \dot{\vartheta}^2)^2}\] \,,
\end{align}
where $V' = \frac{dV}{d\vartheta}$. 

To generate new phenomenology that differs from standard slow-roll inflation (in the absence of the Chern-Simons term), we therefore require
\begin{equation}\label{constraint on oom of alpha eqn}
    \alpha \gtrsim \frac{\Mpl^3}{\dot{\vartheta}\sqrt{V}}\,.
\end{equation}
Using $\epsilon_{\dot\vartheta} = \dot{\vartheta}^2/(2 M_{\rm Pl}^2 H^2)$ and $M_{\rm Pl}^2 H^2 \sim V$, this becomes
\begin{equation}\label{eqn:CScond}
  \alpha \gtrsim   \frac{\Mpl^3}{(\dot{\vartheta}\sqrt{V})} \sim
  \frac{\Mpl^2}{H^2} \frac{1}{\sqrt{\epsilon_{\dot{\vartheta}}} } M_{\rm Pl}^{-1},
\end{equation}
where we have ignored some $\mathcal{O}(10)$ numerical factors. From the constraint on the amplitude of primordial B-modes \cite{Planck:2018jri}, $H/\Mpl < 10^{-5}$, and from the necessity of having a slow-roll background $\sqrt{\epsilon_{\dot{\vartheta}} } < 1$, we find that in order for the Chern-Simons interaction to be important relative to the usual Einstein term, one must have 
\begin{equation}\label{constraint on oom of alpha value}
    \alpha \gtrsim 10^{10} M_{\rm Pl}^{-1}\,.
\end{equation}
In order for the action above to behave as a well-defined effective field theory  (EFT) at the background level, we require it to be valid up to a cutoff of at least $\Lambda \sim H$, which appears to be in conflict with the condition derived from eq.\ \eqref{eqn:CScond}. As we show below in section \ref{subsubsection tensor perturbation chern simons}, the presence of the Chern-Simons term results in the familiar ghost  \cite{Lyth:2005jf,Dyda:2012rj} as well as new gradient instabilities in the spectrum of graviton fluctuations about the inflationary background render questions about the validity of such an effective field theory moot. 

\subsection{Pseudoscalar inflation in Nieh-Yan modified gravity}\label{section eom nieh yan}

Substituting the torsion ansatz into the equations of motion in eq.\ \eqref{einstein equation CS and NY}, and setting $\alpha = 0$ leads to
\begin{align}\label{eom NY}
(H+h)^2 - \phi^2 = & \frac{1}{3\Mpl^2} \left(\frac{\dot{\vartheta}^2}{2} + V(\vartheta)\right),
\\
2 [\dot{H} + \dot{h} + H(H + h)] + (H + h)^2 - \phi^2 + M_{Pl}^{-2}4nf\dot{\vartheta}\phi = & -\frac{1}{\Mpl^2} \left( \frac{\dot{\vartheta}^2}{2} - V(\vartheta) \right),
\\
h = & 0,
\\
 \phi = & \frac{nf\dot{\vartheta}}{\Mpl^{2}}\,,
\end{align}
along with the Klein-Gordon equation for the pseudoscalar, eq. \eqref{KG equation CS and NY} 
\begin{equation}\label{KG equation NY}
\ddot{\vartheta} + \frac{dV}{d\vartheta} + 3H\dot{\vartheta} + 6nf[\dot{\phi} + 3\phi H] = 0\,.
\end{equation}

Unlike the Chern-Simons case, only one torsion field ($\phi$) is nontrivial  in Nieh-Yan modified gravity, as can be seen from eq.\ \eqref{eom NY} (the torsion constraint sets $h = 0$). Moreover, from eq. \eqref{eom CS} and \eqref{eom NY}, if $\vartheta$ is a constant, then the torsion fields vanish and our equations reduce to the usual Friedmann and Klein-Gordon equations. The Klein-Gordon equation \eqref{KG equation NY}, has an extra friction term that is contributed by the torsion fields. Using the expressions for $h$ and $\phi$ in eq.\ \eqref{eom NY}, eq.\ \eqref{KG equation NY} can be written as
 \begin{equation}\label{KG equation NY rewritten}
     \ddot{\vartheta} + 3H\dot{\vartheta} + \frac{1}{\(1 + \frac{6n^2f^2}{\Mpl^2}\)}\frac{dV}{d\vartheta} = 0\,.
 \end{equation}
From eq.\ \eqref{KG equation NY rewritten}, we can see that for Nieh-Yan modified gravity, torsion effectively plays the role of flattening the potential.

Note that the form of eq.\ \eqref{KG equation NY} suggests that one might be able to simply integrate out the torsion field, leading to a modified kinetic term for the scalar field,
\begin{align}
\partial_\mu \vartheta \partial_\nu \vartheta \to \(1 + \frac{6n^2f^2}{\Mpl^2}\)\partial_\mu \vartheta \partial_\nu \vartheta\, ;
\end{align}
where, recall $n = M^2/f^2$. However, as we demonstrate in the next section, this does not hold once (tensor) perturbations about the background are taken into account.

\section{Cosmological perturbations}\label{section perturbations}

Having studied the background solutions in the presence of these pseudo-topological terms and torsion, we turn to the fluctuations about the background to determine the spectra of perturbations. At linear order, the scalar, vector and tensor perturbations do not mix with each other even in the presence of torsion.  We begin by focusing on the tensor fluctuations before computing the scalar fluctuations. The reason for proceeding in this order is that we anticipate pathologies in Chern-Simons modified gravity. In the absence of torsion, it is well known that the theory possesses a ghost instability \cite{Lyth:2005jf,Dyda:2012rj}. In this section, we show that this ghost instability,  together with the emergence of new gradient instabilities, render the theory (with an active Chern-Simons term) inviable.

\subsection{Tensor perturbations} \label{subsection tensor perturbation}

The perturbed metric, around the quasi-de Sitter background, can be written as:
\begin{equation}
    g_{\mu\nu} = \bar{g}_{\mu\nu} + h_{\mu\nu}\,,
\end{equation}
where $h_{\mu\nu}$ is the tensor perturbation which takes the form
\begin{equation}\label{hij tensor fluctuation}
h_{ij} = a^{2}\left[ \delta_{ij} + \gamma_{ij} + \frac{1}{2}\gamma_{im}\gamma_{mj}+\ldots \right]
\end{equation}
with all other components, namely, $h_{00}$ and $h_{0i}$,  zero. The tensor modes are transverse-traceless, $\delta^{ij}\gamma_{ij} = 0$ and $\gamma_{ij,j} = 0$. In terms of tetrads, the tensor fluctuations can be written
\begin{equation}\label{vielbein tensor fluctuation}
e^{a}_{i} = a\Big[\delta^{a}_{i} + \frac{1}{2}\delta^{ak}\gamma_{ki} + \frac{1}{8}\delta^{ak}\gamma_{kj}\gamma_{ji}+\ldots \Big]\,.
\end{equation}
The torsionless part of the spin connection can be written in terms of the tetrads as
\begin{equation}\label{spin connection and vielbein}
    \bar{\omega}_{\nu}^{AB} = e^{A}_{\mu}(\partial_{\nu}e^{B\mu} + \Gamma^{\mu}\hspace{0.3pt}_{\sigma\nu}e^{B\sigma})\,,
\end{equation}
where $\Gamma^{\mu}\hspace{0.3pt}_{\sigma\nu}$ are the usual Christoffel symbols obtained from the metric $g_{\mu\nu}$. The components of the torsionless spin connection, up to quadratic order in the tensor modes, are given by  \cite{Adshead_2019}
\begin{align}\nonumber\label{torsionless spin connection components for tensor perturbation}
\bar{\omega}_{0}^{0b} =  & 0\,,
\\\nonumber
\bar{\omega}_{0}^{ab} = & -\frac{a}{8}\eta^{ac}\eta^{bd}(\dot{\gamma}_{cj}\gamma_{jd} - \gamma_{cj}\dot{\gamma}_{jd})\,,
\\\nonumber
\bar{\omega}_{i}^{0b} = & aH\eta^{bc}\left( \delta_{ci} + \frac{1}{2}\gamma_{ci} + \frac{1}{8}\gamma_{ci}^{2} \right) + \frac{a}{2}\eta^{bc} \left( \dot{\gamma}_{ic} + \frac{1}{2}\gamma _{in}\dot{\gamma}_{nc} \right)\,,
\\\nonumber
\bar{\omega}_{i}^{ab} = & -\eta^{ac}\eta^{bd}\Big[ \frac{1}{2}(\partial_{d}\gamma_{ic} - \partial_{c}\gamma_{id}) + \frac{1}{8}(\gamma_{dk}\partial_{i}\gamma_{kc} - \gamma_{ck}\partial_{i}\gamma_{kd}) + \frac{1}{4}(\partial_{d}\gamma_{ic}^{2} - \partial_{c}\gamma_{id}^{2})\,
\\
 &+ \frac{1}{4}(\gamma_{dk}\partial_{c}\gamma_{ki} - \gamma_{cm}\partial_{d}\gamma_{mi}) + \frac{1}{4}(\gamma_{ck}\partial_{k}\gamma_{id} - \gamma_{dk}\partial_{k}\gamma_{ic}) \Big]\,.
\end{align}
The components of the spin connection, associated with the torsional part, take the form
\begin{align}\label{torsionfull spin connection components for tensor perturbation}
\tilde{\omega}_{i}^{ab} =  \phi \epsilon^{ab}\hspace{0.5pt}_{c} e^{c}_{i}\,, \quad 
\tilde{\omega}^{0b}_{i} = he^{b}_{i}\,, \quad
\tilde{\omega}_{i}^{ab} =  0 \,, \quad
\tilde{\omega}^{0b}_{i} = 0\,.
\end{align}
Next, we use these solutions to find the quadratic action for the primordial gravitational waves (tensor modes) in the presence of the Chern-Simons and Nieh-Yan terms. While both the tensor modes evolve identically in General Relativity, we demonstrate that the parity-violating torsion fields sourced by the Chern-Simons and the Nieh-Yan terms lead to chiral primordial gravitational waves. That is, the left- and right-helical modes satisfy different equations, leading to different mode amplitudes.

\subsubsection{Tensor perturbations in dynamical Chern-Simons modified gravity}\label{subsubsection tensor perturbation chern simons}

It is well-known that the presence of the Pontryagin density during pseudoscalar driven inflation leads to birefringent gravitational waves \cite{Alexander:2004us}. However,  this term is also known to introduce ghost instabilities \cite{Lyth:2005jf,Dyda:2012rj}. With a dynamical axion  $\vartheta$, the kinetic term of one gravitational wave helicity becomes negative in when its physical momenta $k_{\text{phys}} \equiv k/a$, becomes larger than the Chern-Simons mass; $m_{CS} \equiv \Mpl^2/\alpha\dot{\vartheta}$. Our model requires a very large value of the coupling $\alpha$, which implies a very small Chern-Simons mass, and the appearance of the ghost at a low scale. In this section, we first analyze whether the introduction of torsion can resolve this problem.

Substituting eqs.\ \eqref{vielbein tensor fluctuation}, \eqref{torsionless spin connection components for tensor perturbation}, and \eqref{torsionfull spin connection components for tensor perturbation} in eq.\ \eqref{EH action equation} results in the quadratic action for the tensor modes arising from the Einstein-Cartan term
\begin{align}\label{tensor perturbation real space EH action}
S_{\rm EC} 
 = \frac{\Mpl^{2}}{8} \int dt d^{3}x \ a^{3}\Big[\dot{\gamma}_{ba}\dot{\gamma}_{ba}-\frac{1}{a^{2}}\partial_{a}\gamma_{bs}\partial_{a}\gamma_{bs} -   4\frac{\phi}{a}\epsilon^{abc}\gamma_{ap}\partial_{b}\gamma_{pc}\Big]\,.
\end{align}
We expand the tensor modes in Fourier space as
\begin{align}\label{tensor expansion Fourier space}
    \gamma_{ij}(\mathbf{x},\eta) = \sum_{\lambda = \pm 1} \int \frac{d^3 k}{(2\pi)^{3/2}} \gamma^{\lambda}_{\mathbf{k}}(\eta)\epsilon_{ij}^{\lambda}(\mathbf{k})e^{i\mathbf{k}\cdot\mathbf{x}}\,,
\end{align}
where the sum is over the two helicities of  gravitational waves ($\lambda = \pm 1$), and the polarization tensors satisfy
\begin{align}\label{helical polarization tensors relation}
k_{i}\varepsilon^{\pm}_{ij}(\mathbf{k}) = 0 \,, \quad \varepsilon^{\pm}_{ii}(\mathbf{k}) = 0 \,, \quad \epsilon^{ijk} k_{j}\varepsilon^{\pm}_{kl}(\mathbf{k}) = \mp ik\varepsilon^{\pm}_{il}(\mathbf{k}) \,, \nonumber \\
\varepsilon^{\pm}_{ij}(-\mathbf{k}) = \varepsilon^{\mp}_{ij}(\mathbf{k})\hspace{0.5pt}^* \,, \quad \varepsilon^{\lambda}_{ij}(\mathbf{k})\varepsilon^{\lambda'}_{ij}(-\mathbf{k}) = \delta^{\lambda\lambda'}\,.
\end{align}
In Fourier space, eq.\ \eqref{tensor perturbation real space EH action} takes the form
\begin{align}\label{tensor perturbation fourier space EH action}
S_{\rm EC} = \sum_{\lambda} \frac{\Mpl^{2}}{8}\int dt \frac{d^{3}k}{(2\pi)^{3}} a^{3} \Big[ \partial_{0}\gamma_{\mathbf{k}}^{\lambda}\partial_{0}\gamma_{-\mathbf{k}}^{\lambda} -\frac{k}{a} \(\frac{k}{a} + 4\phi\lambda\) \gamma_{\mathbf{k}}^{\lambda}\gamma_{-\mathbf{k}}^{\lambda} \hspace{1pt}  \Big]\,,
\end{align}
while the Chern-Simons term contributes
\begin{align}\label{tensor perturbation fourier space CS action}
S_{\rm CS} 
 = \frac{\alpha}{8}\sum_{\lambda}\int dt \frac{d^{3}k}{(2\pi)^{3}} a^{3} \dot{\vartheta}\[ \(\lambda \frac{k}{a}-\phi\)\partial_{0}\gamma_{\mathbf{k}}^{\lambda}\partial_{0}\gamma_{-\mathbf{k}}^{\lambda} - \frac{k}{a}\(  \lambda \frac{k^{2}}{a^{2}} +2\phi \frac{k}{a}+ \frac{\lambda\phi^{2}}{3}\)\gamma_{\mathbf{k}}^{\lambda}\gamma_{-\mathbf{k}}^{\lambda}\]\,.
\end{align}
The full quadratic action for $\gamma$, including  the Chern-Simons term, is then given by
\begin{align}\label{chern simons total tensor action fourier space}
S^{\rm EC+CS}_{[\gamma^{2}]} = & \sum_{\lambda}\int \frac{\Mpl^{2}}{8} dt \frac{d^{3}k}{(2\pi)^{3}} a^{3} \[ \partial_{0}\gamma_{\mathbf{k}}^{\lambda}\partial_{0}\gamma_{-\mathbf{k}}^{\lambda} -\frac{k}{a} \(\frac{k}{a} + 4\phi\lambda  \) \gamma_{\mathbf{k}}^{\lambda}\gamma_{-\mathbf{k}}^{\lambda} \hspace{1pt}  \]
\\\nonumber
& + \frac{\alpha}{8}\sum_{\lambda}\int dt \frac{d^{3}k}{(2\pi)^{3}} a^{3} \dot{\vartheta}\[ \( \lambda \frac{k}{a}-\phi \)\partial_{0}\gamma_{\mathbf{k}}^{\lambda}\partial_{0}\gamma_{-\mathbf{k}}^{\lambda} -\frac{k}{a} \( \lambda \frac{k^{2}}{a^{2}} +2\phi \frac{k}{a} + \frac{\lambda\phi^{2}}{3}\)\gamma_{\mathbf{k}}^{\lambda}\gamma_{-\mathbf{k}}^{\lambda}\]\,.
\end{align}
With the quadratic action in hand, we can read off the dispersion relation
\begin{align}
\omega^2 = \frac{\Mpl^2\Big(\frac{k}{a} + 4\phi\lambda   \Big)\frac{k}{a}  + \alpha \dot{\vartheta}\Big(\lambda  \frac{k^{2}}{a^{2}} +2\phi \frac{k}{a} +\frac{\lambda\phi^{2}}{3}\Big)\frac{k}{a}}{\Mpl^2 + \alpha \dot{\vartheta}\left( \lambda \frac{k}{a}-\phi\right)}.
\end{align}
The theory remains ghost-free if, for the momenta of interest, 
\begin{align}
\Mpl^2 + \alpha \dot{\vartheta}\left( \lambda \frac{k}{a}-\phi\right) > 0,
\end{align}
substituting in 
\begin{align}\label{phiconstr}
\phi = \frac{\alpha \dot{\vartheta}}{6M_{\rm Pl}^4}\rho ,
\end{align}
using the estimate from eq.\ \eqref{constraint on oom of alpha eqn}, and approximating $\rho \sim V$, we can write this as a restriction on momenta for the unstable helicity
\begin{align}
\frac{k}{a} \lesssim \frac{1}{2}\frac{\sqrt{V}}{M_{\rm Pl}}  \sim H.
\end{align}
This indicates that the model has a ghost that appears for scales that are near the horizon. Ignoring the validity of the EFT, our theory is unstable for all modes that are inside the horizon.

We also note that the presence of torsion leads to a potential gradient instability. Stability also requires 
\begin{align}
\Mpl^2\frac{k}{a}\Big(\frac{k}{a} + 4\phi\lambda  \Big)  + \alpha  \dot{\vartheta}\frac{k}{a}\Big(\frac{\lambda\phi^{2}}{3}+ \lambda \frac{k^{2}}{a^{2}} +2\phi\frac{k}{a}  \Big) > 0
\end{align}
which we can estimate using eq.\ \eqref{phiconstr}, for $k \neq 0$,
\begin{align}
2\frac{k}{a}   +\lambda\frac{\Mpl}{\sqrt{V}}\(\frac{ k^{2}}{a^{2}}+ \frac{25}{24}\frac{\sqrt{V}}{\Mpl}\) > 0
\end{align}
and we see that at both large and small wave numbers, one of the helicities appears to have the wrong sign for its gradient term. Given these pathologies, we do not consider inflation in dynamical Chern-Simons gravity (with the coupling large enough to make the Chern-Simons interaction relevant for the background solution) to be viable.

%
\subsubsection{Tensor perturbations in Nieh-Yan modified gravity}\label{subsubsection tensor perturbation Nieh Yan}
%

We next consider the tensor fluctuations when the Einstein-Hilbert action is supplemented by the Nieh-Yan term supported by torsion. When the tensor field is turned on, the Nieh-Yan action at quadratic order is given  in Fourier space  by (with the same Fourier conventions as eq.\ \eqref{tensor expansion Fourier space})
\begin{align}\label{tensor perturbation fourier space NY action}
S_{\rm NY} =  \sum_{\lambda = \pm 1}\lambda nf \int dt \frac{d^{3}k}{(2\pi)^{3}}a^{2} \dot{\vartheta} k \gamma^{\lambda}_{\mathbf{k}} \gamma^{\lambda}_{-\mathbf{k}}\,.
\end{align}
In this case, the full quadratic action for  $\gamma^\lambda$ is
\begin{align}\label{nieh yan total tensor action fourier space}
S^{\rm EC+NY}_{[\gamma^{2}]} 
= \sum_{\lambda = \pm 1}\int dt \frac{d^{3}k}{(2\pi)^3}  \frac{M_{Pl}^2 a^3}{8}\Big[ \partial_{0}\gamma_{\mathbf{k}}^{\lambda}\partial_{0}\gamma_{-\mathbf{k}}^{\lambda} - \Big(\frac{k^2}{a^{2}} - \frac{4\lambda\phi k}{a}  \Big) \gamma_{\mathbf{k}}^{\lambda}\gamma_{-\mathbf{k}}^{\lambda}  \Big]\,.
\end{align}
Varying the action gives us the following equations of motion for the tensor modes:
\begin{align}\label{nieh yan tensor eqn of motion normal time}
\frac{\partial^2 \gamma_{\mathbf{k}}^{\pm}}{\partial t^2} + 3H\frac{\partial \gamma_{\mathbf{k}}^{\pm}}{\partial t} + \Big(\frac{k^2}{a^{2}} \mp \frac{4\phi k}{a}  \Big) \gamma_{\mathbf{k}}^{\pm} = 0\,,
\end{align}
where note that the left and right helicity modes obey different equations of motion in the presence of torsion.

Working in conformal time
\begin{align}
\eta =\int^t_{-\infty} \frac{dt}{a(t)}, 
\end{align}
we obtain
\begin{align}\label{nieh yan tensor eqn of motion conformal time}
\frac{\partial^2 \gamma_{\mathbf{k}}^{\pm}}{\partial \eta^2} + 2\mathcal{H} \frac{\partial \gamma_{\mathbf{k}}^{\pm}}{\partial \eta} + \Big(k^2 \mp 4\phi ak  \Big) \gamma_{\mathbf{k}}^{\pm} = 0\,,
\end{align}
where $\mathcal{H} = a'/a$ is the conformal Hubble parameter and $'$ represents a derivative with respect to conformal time.

%
\subsection{Tensor power spectrum}\label{subsection tensor power spectrum}
%

Ignoring the contributions due to the Pontryagin density, we solve eq.\ \eqref{nieh yan tensor eqn of motion conformal time} with Bunch-Davies vacuum conditions to obtain the resulting gravitational wave spectrum at horizon crossing. To obtain an analytic solution, we approximate $\dot{\vartheta}$ and $H$ as constants. Note that, through the constraint,  this condition implies that the torsion $\phi$ field is also constant during the slow-roll phase. We canonically normalize the fields by redefining
\begin{align}
\gamma_{\mathbf{k}}^{\pm} = \frac{2 }{a \Mpl}u_{\mathbf{k}}^{\pm}.
\end{align}
Working with $x \equiv k\eta$, the equation of motion for the gravitational waves,  eq.\ \eqref{nieh yan tensor eqn of motion conformal time}, reduces to
\begin{align}\label{nieh yan tensor eqn of motion conformal time with approximation}
\frac{\partial^2 u_{\mathbf{k}}^{\pm}}{\partial x^2} + \(1 \mp \frac{4\phi}{Hx} - \frac{2 + 3\epsilon_H}{x^2}\)u_{\mathbf{k}}^{\pm} = 0\,.
\end{align}
The general solutions to eq.\ \eqref{nieh yan tensor eqn of motion conformal time with approximation} are the Coulomb wave functions or the Coulomb-Hankel function, $H^{\pm}_{l}(\beta,\rho)$. In our case, $\rho = x$, $l = 1 + \epsilon_H$ and $\beta = \pm \frac{2\phi}{H}$. Imposing Bunch-Davies initial conditions as $x \rightarrow -\infty$ selects the $H_{l}^{-}(\beta,\rho)$ solution, while canonical quantization fixes its amplitude. At horizon crossing, the solutions for the canonically normalized tensor mode functions are (see appendix \ref{Appendix Expression for Tensor power spectrum})
\begin{align}\label{transformed tensor mode function solutions}
|u^{\pm}_{\mathbf{k}}|^2 = \frac{1}{2k}\[\frac{(k\eta)^{-2}}{\(3C_{1 + \epsilon_H}(\pm \frac{2\phi}{H})\)^2}\]\,,
\end{align}
where 
\begin{align}
C_{l}(\beta) = \frac{2^{l}e^{-\pi \beta/2}|\Gamma(l+1+i\beta)|}{(2l+1)!}.
\end{align} 
The tensor power spectrum is given by
\begin{align}\label{plus minus tensor power spectrum formula}
    \mathcal{P}_{T}^{\pm}(k) = \frac{k^3}{2\pi^2}\left| \frac{2u_{\mathbf{k}}^{\pm}}{\Mpl a}\right|^2\,,
\end{align}
which results in
\begin{align}\label{plus minus tensor power spectrum NY}
\mathcal{P}_{T}^{\pm}(k) = \frac{H^2}{\Mpl^2\pi^2} \frac{e^{\pm 2\pi\phi/H}}{|\Gamma(2 + \epsilon_H \pm 2\phi i/H)|^{2}}\,.
\end{align}
The total tensor power spectrum is the sum over the individual helicities
\begin{align}\label{total tensor power spectrum formula}
        \mathcal{P}_{T}(k) =  \mathcal{P}_{T}^{+}(k) + \mathcal{P}_{T}^{-}(k)\,.
\end{align}
Using eq.\ \eqref{plus minus tensor power spectrum NY}, the expression for the total power spectrum then takes the form
\begin{align}\label{eqn total tensor power spectrum nieh yan}
        \mathcal{P}_{T}(k) = \frac{H^2}{\Mpl^2\pi^3}\frac{\sinh\Big(\frac{4\phi\pi}{H}\Big)}{\Big(1 + \frac{4\phi^2}{H^2}\Big)\Big(\frac{2\phi}{H}\Big)}\,.
\end{align}
where we have used the approximation $\epsilon_H < 1$ during inflation and hence, can be ignored for $|\Gamma(2 + \epsilon_H \pm 2\phi i/H)|^{2}$. In the torsionless limit $\phi \rightarrow 0$, eq.\ \eqref{eqn total tensor power spectrum nieh yan} reduces to the familiar expression for the dimensionless tensor power spectrum 
\begin{align}
\mathcal{P}_{T}(k) = \frac{2H^2}{\Mpl^2\pi^2}\,.
\end{align}

To understand the modifications to the spectrum coming from torsion, note that the slow roll parameter for Nieh-Yan modified gravity, in terms of the Hubble flow, is given by
\begin{align}\label{eqn slow roll nieh yan}
\epsilon_{H} = \frac{\phi^2}{2n^2f^2H^2} + \frac{3\phi^2}{H^2}\,.
\end{align}
The total tensor power spectrum is exponentially enhanced for large values of $|\phi/H|$. However, for slow-roll trajectories, $\epsilon_{H} < 1$ constrains $\phi^2/H^2 < 1/(3 + (\sqrt{2}nf)^{-2}) < 1$. The amplitude of the total tensor power spectrum, therefore, does not get a significant enhancement during inflation in the presence of the Nieh-Yan density. 

Nevertheless, an order unity $|\phi/H|$ can lead to a large chirality,  defined by
\begin{equation}\label{chirality formula}
    \chi = \left|\frac{\mathcal{P}_{T}^{+}(k)-\mathcal{P}_{T}^{-}(k)}{\mathcal{P}_{T}^{+}(k)+\mathcal{P}_{T}^{-}(k)}\right|\,.
\end{equation}
Substituting eq.\ \eqref{plus minus tensor power spectrum NY}, the chirality parameter is given by
\begin{equation}
    \chi = \left|\tanh\( \frac{4\pi\phi}{H}\)\right|.
\end{equation}
While chiral gravitational waves provide an interesting observational signature, we note that they frequently arise in models of inflation driven by pseudoscalars coupled to gauge and fermion fields \cite{Barnaby:2011qe, Maleknejad:2011jw, Maleknejad:2011sq,  Anber:2012du, Dimastrogiovanni:2012ew, Maleknejad:2012fw, Adshead:2013nka, Adshead:2013qp, Pajer:2013fsa, Namba:2013kia,Maleknejad:2014wsa, Adshead:2016omu, Dimastrogiovanni:2016fuu, Maleknejad:2016qjz, Adshead:2017hnc,  Adshead:2018doq, Dimastrogiovanni:2018xnn, Adshead:2019aac, Okano:2020uyr,Weiner:2020sxn, Iarygina:2021bxq, Dimastrogiovanni:2023oid, Michelotti:2024bbc, Aoki:2025uwz}.

So far, we have been largely agnostic about the form of the potential for the pseudoscalar field. In order to get some intuition about how large the chirality is for the Nieh-Yan density, and to check the consistency of our analytical solutions, we introduce the cosine potential for natural inflation (CNI) \cite{PhysRevLett.65.3233}, given by
\begin{equation}\label{cosine natural inflation}
V(\vartheta) = \Lambda^4\[1-\cos\(\frac{\vartheta}{f}\)\]\,,
\end{equation}
where $\Lambda$ is an energy scale.

\begin{figure}[t]
   \includegraphics[width = 1.01\linewidth]{"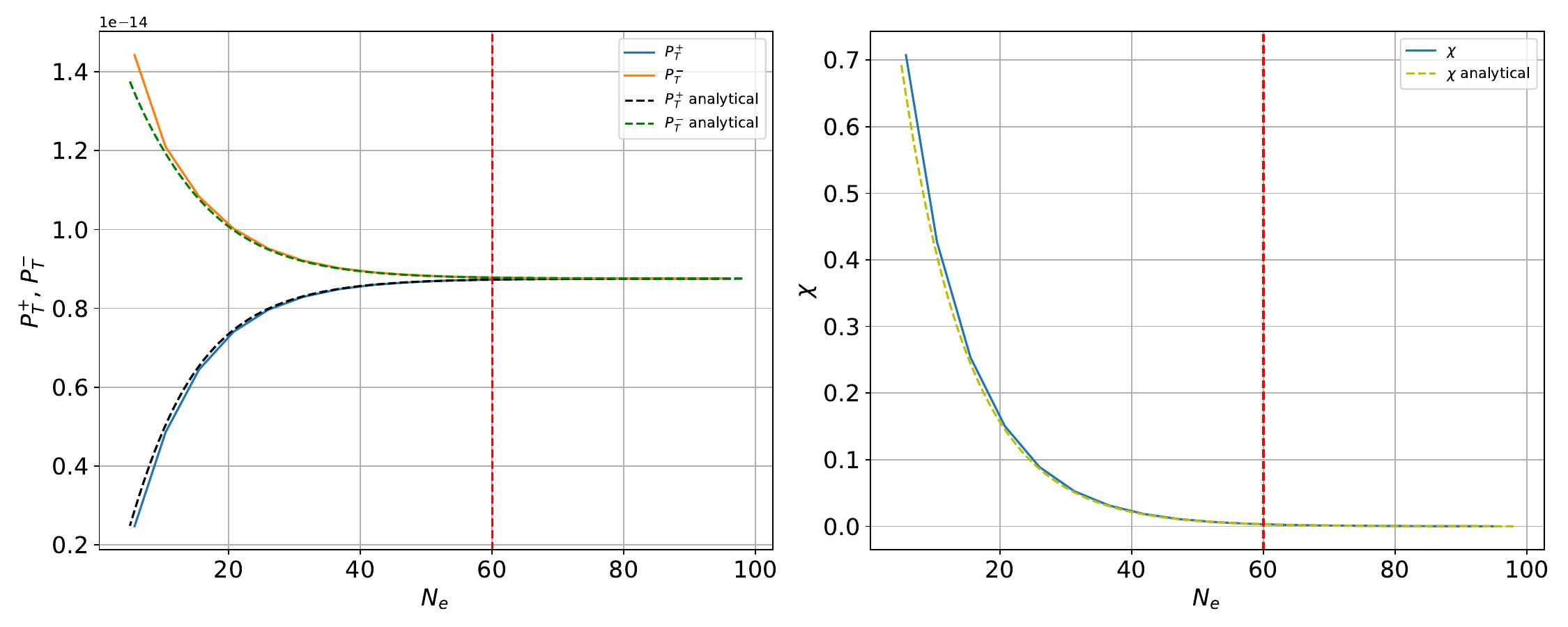"}  
  \caption{Tensor powerspectrum, $P_{T}$, and chirality parameter, $\chi$ as a function of $e$-folding number for  natural inflation (eq.\ \eqref{cosine natural inflation}). Here, $N_e$ is the number of $e$-folds before inflation ends, $f/\Mpl  = 0.1$, $n = 90$, $\Lambda = 0.6\times10^{-3}\Mpl $.}
  \label{fig Pt and chi f point 2 n 65 lambda 5 point 9 10 minus 3 cni}
\end{figure}

In figure \ref{fig Pt and chi f point 2 n 65 lambda 5 point 9 10 minus 3 cni}, we show the tensor power spectrum and resulting chirality parameter for natural inflation  (eq.\ \ref{cosine natural inflation}) with $n = 90$ and $f/\Mpl  = 0.1$. $\Lambda = 0.6\times10^{-3}\Mpl$. The parameters are chosen such that the amplitude of the scalar power spectrum is fixed to approximately match the Planck value (see below),  $A_{s} (N_e = 60) \simeq 2.1\times 10^{-9}$. We observe from the plot that the analytical expressions are excellent approximations of the numerical solutions.

\begin{figure}[t]
  \centering
  \includegraphics[width=\linewidth]{"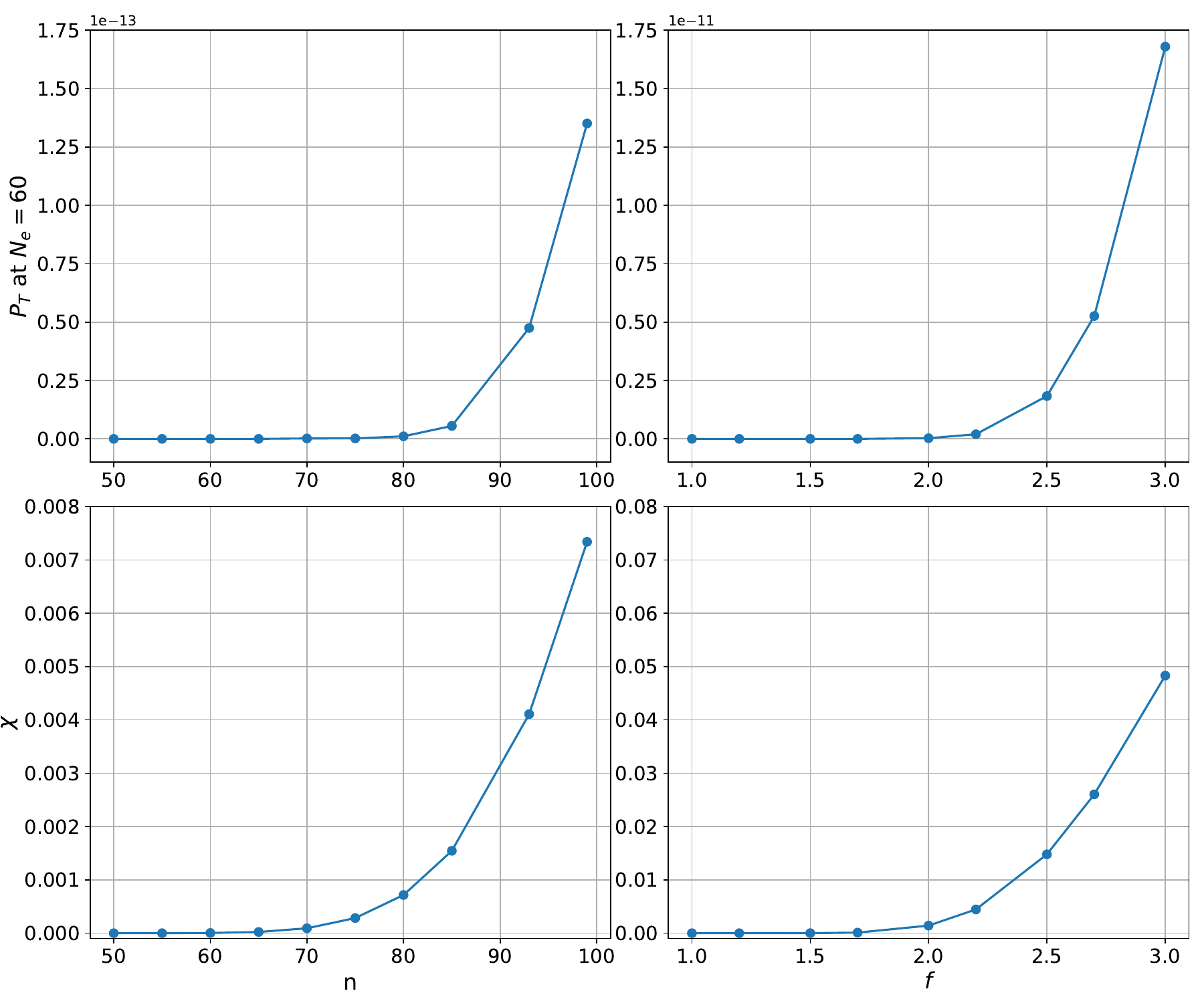"}
  \caption{(Left panel) Tensor powerspectrum, $P_{T}$, and chirality parameter, $\chi$, at $N_e = 60$ $e$-folds before the end of inflation as $n$ is varied, $n \hspace{2pt} \in \hspace{2pt} \{50,99\}$, and fixed axion decay constant $f/\Mpl = 0.1$. (Right panel) Tensor powerspectrum, $P_{T}$, and chirality parameter, $\chi$, as the axion decay constant  $f$ is varied, $f/\Mpl  \hspace{2pt} \in \hspace{2pt} \{1,3\}$, with a fixed $n= 0.1$. We fix the energy scale, $\Lambda = 10^{-8}-10^{-3} \Mpl $, such that the scalar amplitude $A_{s} (N_e = 60) \simeq 2.1\times 10^{-9}$ (see section \ref{results}). For both panels, the potential is given by  eq.\ \eqref{cosine natural inflation}. }
  \label{fig tensor amplitude and chi vs n and f nieh yan}
\end{figure}

From figure \ref{fig tensor amplitude and chi vs n and f nieh yan}, we see that the magnitude of the tensor amplitude and chirality at $60$ $e$-folds before inflation ends, keeps increasing with increasing $n$ with a fixed value of $f$ (or vice versa).

\subsection{Scalar perturbations}\label{subsection scalar perturbation in Nieh Yan}
We work in the spatially flat gauge, where the metric is given by
\begin{equation}\label{metric scalar perturbation}
[g_{\mu\nu}] = a^{2}(\eta) [\eta_{\mu\nu} + h_{\mu\nu}] =
  a^{2}(\eta)\begin{bmatrix}
   1+2A &
   -\partial_{i}B \\
   -\partial_{i}B &
   -\delta_{ij} 
   \end{bmatrix}\,,
\end{equation}
where $\eta$ is conformal time, as before. Hence, the components of the tetrad field $e^{A}\hspace{0.5pt}_{\mu}$ are given by \cite{Golovnev:2020aon}
\begin{align}\label{vielbein scalar perturbation}
e^{0}\hspace{0.5pt}_{0} = a[1+A] , \quad e^{0}\hspace{0.5pt}_{i} = a\partial_{i}\beta , \quad e^{a}\hspace{0.5pt}_{0} = a\delta^{ai}\partial_{i}\zeta , \quad e^{a}\hspace{0.5pt}_{i} = a[\delta_{ia} + \epsilon_{aik}\partial_{k}s]\,.
\end{align}
where we have defined $B = \zeta - \beta$ and $s$ is a pseudo-scalar. In the scalar sector, we also perturb the torsion and axion field
\begin{align}
h =  h(\eta) + \delta h(\eta,\vec{x})\,,\quad
\phi =  \phi(\eta) + \delta \phi(\eta,\vec{x})\,,\quad
\vartheta =  \vartheta(\eta) + \delta \vartheta(\eta,\vec{x})\,.
\end{align}

After some straightforward but tedious calculations (see appendix \ref{Appendix Equation of motion for the scalar perturbation}), the final equation of motion for $\delta\vartheta$ in Fourier space is given by 
\begin{align}\nonumber\label{nieh yan scalar eqn of motion conformal time}
\delta \vartheta'' + 2\mathcal{H}\vartheta' + k^2 \delta\vartheta + \frac{a^2}{1 + \frac{6n^2f^2}{\Mpl^2}} \frac{d^2V}{d\vartheta^2}\delta\vartheta
- \vartheta'\( 1+ \frac{6n^2f^2}{\Mpl^2}\)\[ \frac{\vartheta''\delta\vartheta}{\mathcal{H}} + \frac{\vartheta'\delta\vartheta'}{\mathcal{H}} - \frac{\vartheta'\delta\vartheta\mathcal{H}'}{\mathcal{H}^2} \]
\\
+ \frac{\vartheta'}{2\mathcal{H}}\[ \(1 + \frac{6n^2f^2}{\Mpl^2}\)\vartheta'\delta\vartheta' + a^2\frac{dV}{d\vartheta}\delta\vartheta + \(1+\frac{6n^2f^2}{\Mpl^2}\)\frac{2\vartheta'\delta\vartheta}{\mathcal{H}}a^2V \] + a^2\frac{dV}{d\vartheta} \frac{\vartheta' \delta\vartheta}{\mathcal{H}} = 0\,,
\end{align}
where a prime, $\; '\; $, represents a derivative with respect to conformal time. We can see from eq.\ \eqref{nieh yan scalar eqn of motion conformal time} that the pseudo-scalar ($s$) plays no role in the equation of motion. Going back to cosmic time, eq.\ \eqref{nieh yan scalar eqn of motion conformal time} reduces to
\begin{align}\nonumber\label{nieh yan scalar eqn of motion cosmic time}
\ddot{\delta\vartheta} + 3H\dot{\delta\vartheta} + \frac{k^2}{a^2}\delta\vartheta + \frac{1}{1 + \frac{6n^2f^2}{\Mpl^2}} \frac{d^2V}{d\vartheta^2}\delta\vartheta -\dot{\vartheta}\(1+\frac{6n^2f^2}{\Mpl^2}\)\[ \frac{\ddot{\vartheta}\delta\vartheta}{H} + \frac{\dot{\vartheta}\dot{\delta\vartheta}}{H} - \frac{\dot{\vartheta}\dot{H}\delta\vartheta}{H^2}  \]
\\
-\frac{\dot{\vartheta}}{2H}\[ \( 1+\frac{6n^2f^2}{\Mpl^2} \)\dot{\vartheta}\dot{\delta\vartheta} + \frac{dV}{d\vartheta}\delta\vartheta + \(1+\frac{6n^2f^2}{\Mpl^2}\)\frac{2\dot{\vartheta}\delta\vartheta}{H}V \] + \frac{\dot{\vartheta}}{H} \frac{dV}{d\vartheta} \delta\vartheta = 0\,.
\end{align}
Here the overdot `$\; \dot{}\; $' represents a derivative with respect to cosmic time, $t$. We perform a variable transformation to go to the canonical field 
\begin{align}
\delta\vartheta = -\frac{v}{a\sqrt{\(1 + \frac{6n^2f^2}{\Mpl^2} \)}}.
\end{align}
Eq.\ \eqref{nieh yan scalar eqn of motion cosmic time} then becomes
\begin{align}\nonumber\label{nieh yan scalar eqn of motion cosmic time redefined}
\ddot{v} = -H\dot{v} + v\Bigg[ -\frac{k^2}{a^2} + \frac{5}{2}\( 1 + \frac{6n^2f^2}{\Mpl^2} \)\dot{\vartheta}^2 + 2H^2 + 2\frac{\dot{\vartheta}\ddot{\vartheta}}{H}\( 1 + \frac{6n^2f^2}{\Mpl^2} \)
\\
+ \frac{\(1 + \frac{6n^2f^2}{\Mpl^2} \)^2}{2}\frac{\dot{\vartheta}^4}{H^2} - \frac{1}{1 + \frac{6n^2f^2}{\Mpl^2}}\frac{d^2V}{d\vartheta^2}\Bigg]\,.
\end{align}

As a sanity check, note that in the limit $n = 0$, this reduces to the usual Mukhanov-Sasaki equation for standard inflation in Einstein gravity
\begin{equation}\label{normal inflation scalar eqn of motion cosmic time redefined}
\ddot{v} = -H\dot{v} + v\[ -\frac{k^2}{a^2} + \frac{5}{2}\dot{\vartheta}^2 + 2H^2 + 2\frac{\dot{\vartheta}\ddot{\vartheta}}{H} + \frac{1}{2}\frac{\dot{\vartheta}^4}{H^2} - \frac{d^2V}{d\vartheta^2}\]\,.
\end{equation}

Comparing eqs.\ \eqref{nieh yan scalar eqn of motion cosmic time} and \eqref{normal inflation scalar eqn of motion cosmic time redefined}, it is immediately apparent that eqs.\ \eqref{nieh yan scalar eqn of motion cosmic time} can be transformed into eq.\ \eqref{normal inflation scalar eqn of motion cosmic time redefined} by redefining $\vartheta$ to $\sqrt{1 + \frac{6n^2f^2}{\Mpl^2}}\,\vartheta$.\footnote{Notice that the canonical normalization already rescaled $v \rightarrow \sqrt{1 + \frac{6n^2f^2}{\Mpl^2}}\,\delta\vartheta$. } This is simply a consequence of the fact that integrating out the torsion field leads to the axion coupled to usual Einstein gravity in the scalar sector. However, as we have seen in the previous section, the tensor sector picks up corrections that cannot be accounted for in this way.

%
\subsection{Scalar power spectrum}\label{subsection scalar power spectrum}
%

In order to find an analytical solution to eq.\ \eqref{nieh yan scalar eqn of motion cosmic time redefined}, we work in the approximation that $\dot{\vartheta}$,  $\phi$, $H$ and $d^2V/d\vartheta^2$, are constant. In conformal time,  the equation of motion for the canonically normalize scalar fluctuations
\begin{equation}\label{appx nieh yan scalar eqn of motion conformal time redefined}
v'' = v\[ -k^2 + (aH)^2 \Big(\frac{5}{2}\(1 + \frac{6n^2f^2}{\Mpl^2} \)\frac{\dot{\vartheta}^2}{H^2} + 2 + \frac{\(1 + \frac{6n^2f^2}{\Mpl^2} \)^2}{2}\frac{\dot{\vartheta}^4}{H^4} - \frac{1}{H^2 \(1 + \frac{6n^2f^2}{\Mpl^2} \)}\frac{d^2V}{d\vartheta^2}\Big) \]\,,
\end{equation}
can be written
\begin{equation}\label{appx nieh yan scalar eqn of motion conformal time rewritten}
\frac{d^2v}{d\eta^2} + \Big[k^2 - \frac{\nu^2 - \frac{1}{4}}{\eta^2} \Big]v = 0\,,
\end{equation}
where 
\begin{align}
a(\eta) \simeq \frac{1+\epsilon_H}{H\eta}
\end{align}
 and 
 \begin{align}\nonumber
\nu^2 - \frac{1}{4} = {\left\{1 + \left(\frac{1 + 6n^2f^2}{\Mpl^2}\right)\frac{\dot{\vartheta}^2}{H^2}\right\}} \[\frac{5}{2}\left( 1 + \frac{6n^2f^2}{\Mpl^2} \right)\frac{\dot{\vartheta}^2}{H^2} + 2 + \frac{\left( 1 + \frac{6n^2f^2}{\Mpl^2} \right)^2}{2}\frac{\dot{\vartheta}^4}{H^4}\right.
\\
 \left.- \frac{1}{H^2 \(1 + \frac{6n^2f^2}{\Mpl^2} \)} \frac{d^2V}{d\vartheta^2} \]
 \end{align}
 is a constant. The general solution to eq.\ \eqref{appx nieh yan scalar eqn of motion conformal time rewritten} can be written in terms of the Hankel functions of the first and second kind, $H_{\nu}^{1}$, and  $H_{\nu}^{2}$. The modes are quantized by imposing the canonical quantization condition, which leads to the Wronskian condition on the modefunctions
\begin{align}\label{eqn:canquan}
[v,v'] = i.
\end{align}
The Bunch-Davis initial conditions as $\eta \rightarrow -\infty$ singles out the $H_{\nu}^{1}$ solution,
which, together with eq.\ \eqref{eqn:canquan}, sets
\begin{equation}\label{NY scalar mode solution}
v(x) = \frac{1}{\sqrt{2k}}\sqrt{\frac{\pi}{2}}e^{i\big(\nu + \frac{1}{2}\big)\frac{\pi}{2}}\sqrt{-x}\,H_{\nu}^{1}(-x)\,,
\end{equation}
where we have defined $x = k\eta$, as before. 
The curvature perturbation in spatially flat gauge is defined as
\begin{equation}\label{curvature perturbation definition}
\mathcal{R} = -\frac{H}{\dot{\vartheta}}\delta\vartheta\,,
\end{equation}
while the curvature power spectrum is given by
\begin{equation}\label{curvature power spectrum definition}
P_{\mathcal{R}} = \frac{k^3}{2\pi^2}|\mathcal{R}|^2 \,.
\end{equation}
In the limit $x \rightarrow 0$, the Hankel function behaves as
\begin{equation}\label{Hankel fn first kind zero asymptotic form}
H_{\nu}^{1}(x) = \sqrt{\frac{2}{\pi}} e^{-i\frac{\pi}{2}}2^{\nu - \frac{3}{2}}\frac{\Gamma(\nu)}{\Gamma\( \frac{3}{2} \)} (x)^{-\nu}\,.
\end{equation}
At horizon crossing, the curvature power spectrum can therefore be evaluated as 
\begin{equation}\label{nieh yan curvature power spectrum eqn}
P_{\mathcal{R}} = \frac{H^2}{4\pi^2\Mpl^2\(1 + \frac{6n^2f^2}{\Mpl^2}\)}\(\frac{H}{\dot{\vartheta}}\)^2 2^{2\nu - 3}\left|\frac{\Gamma(\nu)}{\Gamma\( \frac{3}{2} \)}\right|^2\,.
\end{equation}
where $\Gamma(x)$ is the Gamma function, and the factor of $\(1 + \frac{6n^2f^2}{\Mpl^2}\)^{-1}$ comes from the transformation of the canonical variable $v$ back to $\delta\vartheta$.

\begin{figure}[t]
  \centering
  \includegraphics[width=0.96\linewidth]{"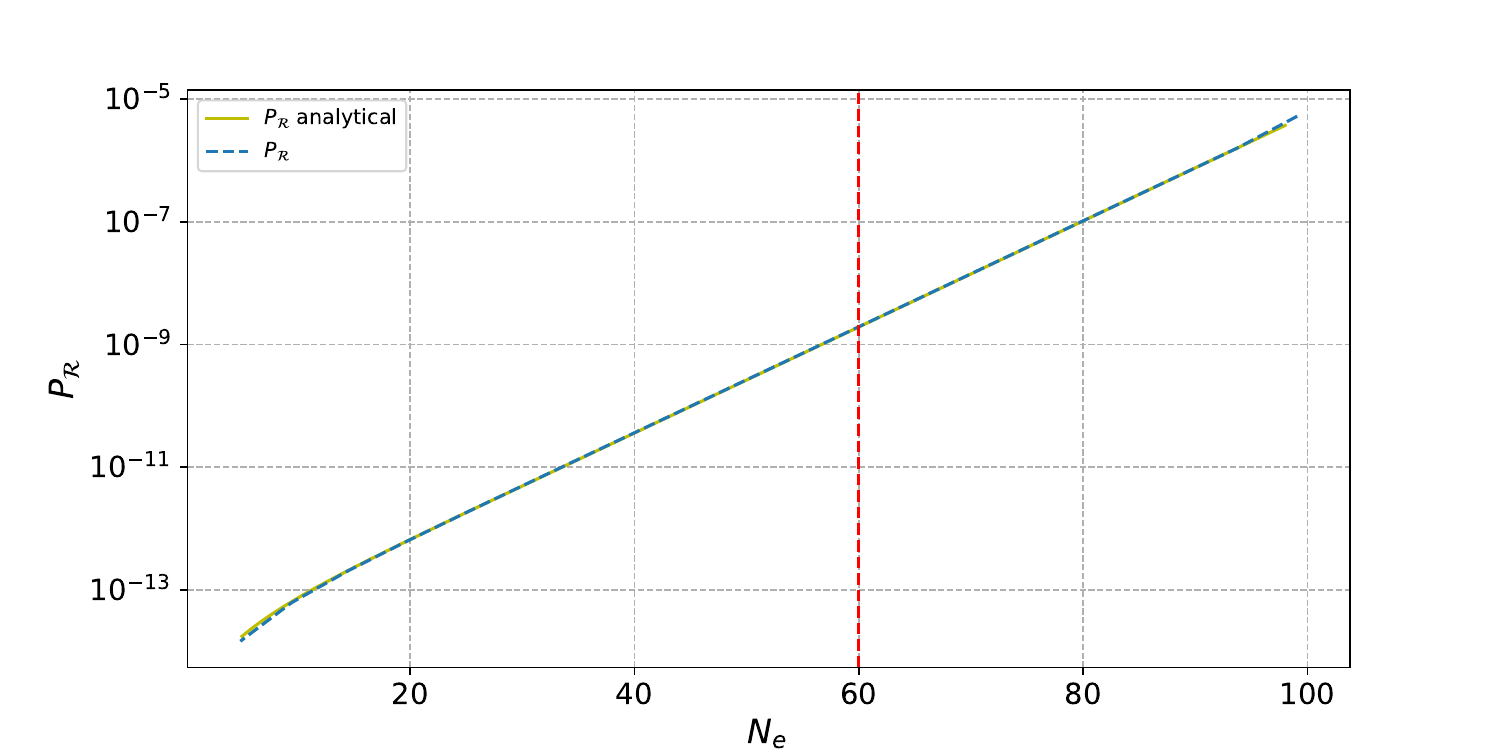"}  
  \caption{$P_{\mathcal{R}}$ vs $N_e$ for CNI model, where $N_e$ is the number of $e$-folds before inflation ends, $\Mpl = 1$, $f = 0.1$, $n = 90$, $\Lambda = 0.6\times10^{-3}$}
  \label{fig Ps f point 2 n 65 lambda 1 point 0391 10 minus 3}
\end{figure}

In figure \ref{fig Ps f point 2 n 65 lambda 1 point 0391 10 minus 3} we show the spectrum of scalar fluctuations for natural inflation with a cosine potential (eq \eqref{cosine natural inflation}) with $n = 90$, $f/\Mpl = 0.1$ and $\Lambda = 0.6\times10^{-3} \Mpl$ note the excellent agreement between the analytical expression and the full numerical solutions.

%
\section{Phenomenology}\label{results}

In order to explore the phenomenological implications of our model, we examine the predictions for various inflationary potentials in the $n_s$-$r$ plane. For illustrative purposes, we consider three distinct models: Generalized Natural Inflation (GNI), Hilltop Squared Inflation (HSI), and D-brane Inflation (DB), and demonstrate how they can be brought into agreement with observational data for sub-Planckian decay constants by coupling to the Nieh-Yan term. 

\begin{figure}[t]
  \centering
  \includegraphics[width=\linewidth]{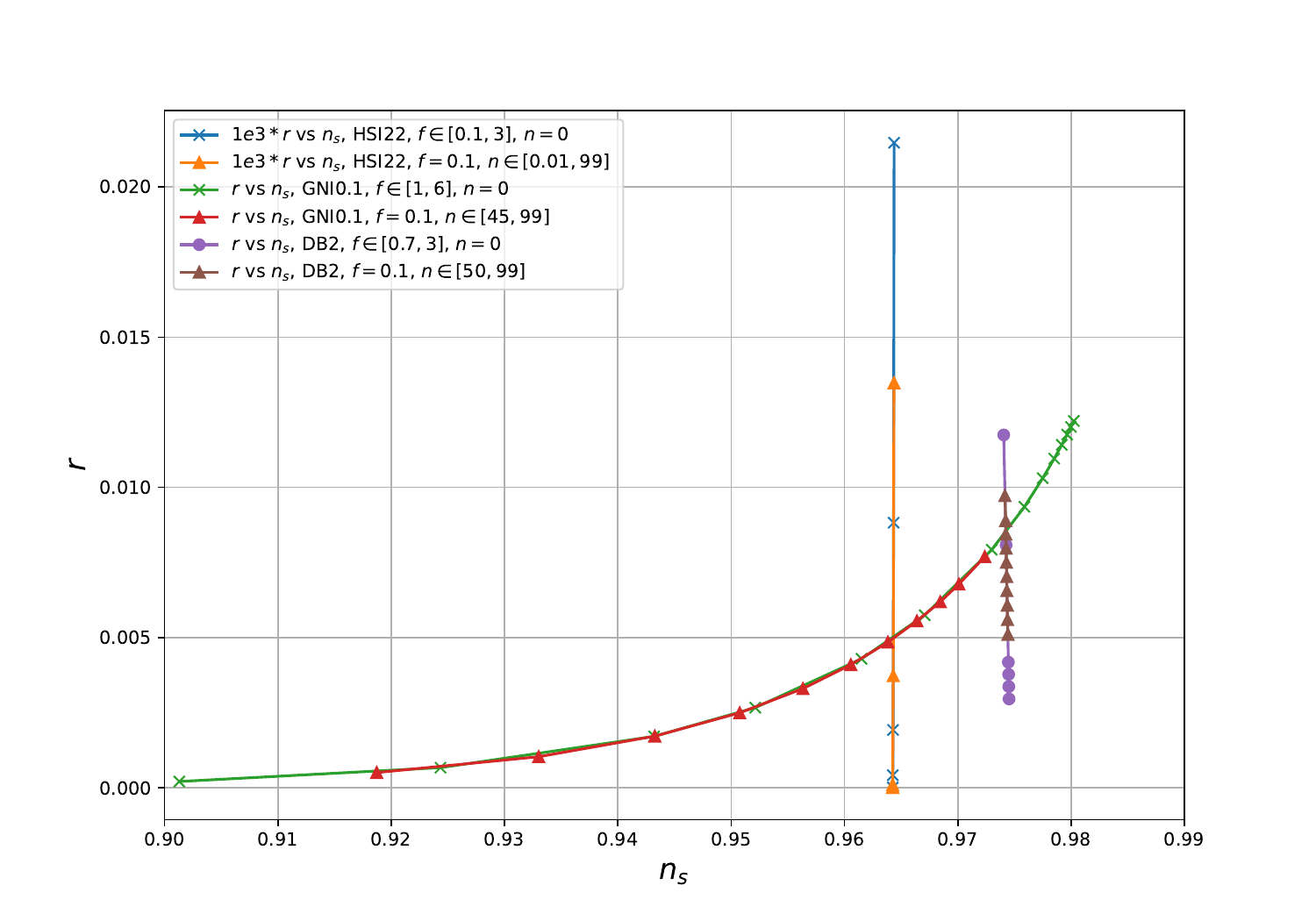}
  \caption{$r$ vs $n_s$ for HSI$22$, GNI$0.1$ and DB$2$ model, $\Mpl = 1$.}
  \label{fig r vs ns sqh cni nieh yan}
\end{figure}

We begin with the \textit{Generalized Natural Inflation} model (labelled ``GNI$0.1$'' in the figures). The potential is given by \cite{zhang2020constraintsgeneralizednaturalinflation, Mu_oz_2015}:
\begin{equation}\label{general natural inflation}
    V(\vartheta) = 2^{1-p}\Lambda^4\left[ 1 + \cos(\vartheta/f)\right]^p \,.
\end{equation}
We adopt the power $p = 0.1$. In figure \ref{fig r vs ns sqh cni nieh yan}, we compare the predictions of this model in standard Einstein gravity ($n=0$) with a variable decay constant ($f\in [0.1, 6]$) against Nieh-Yan modified gravity ($n \neq 0$) for a fixed sub-Planckian decay constant $f=0.1\,\Mpl$. Notably, the curves for GNI in Einstein gravity and Nieh-Yan modified gravity lie nearly on top of each other. This indicates that while the addition of the Nieh-Yan interaction and torsion allows the model to work with much smaller axion decay constants, it largely preserves the correlation between observables found in General Relativity, effectively rescaling the axion decay constant.

Next, we collect the results for \textit{Hilltop Squared Inflation} (HSI) and the \textit{D-brane} model (DB). We define the potentials as follows:
\begin{align}
    V(\vartheta) &= \Lambda^4\left[ 1- \left(\frac{\vartheta}{f}\right)^p\right]^2 \quad (\text{HSI})\,, \\
    V(\vartheta) &= \Lambda^4\left[ 1- \left(\frac{f}{\vartheta}\right)^p\right]^2 \quad (\text{D-brane})\,.
\end{align}
For the HSI model, recent analyses suggest that taking $p = 22$ allows for $n_s \approx 0.965$ \cite{lynker2025actimplicationshilltopinflation}; we label this case ``HSI22''. For the D-brane model, we select $p = 2$ (labeled ``DB2'') following \cite{Adshead_2020}.

As shown in figure \ref{fig r vs ns sqh cni nieh yan}, the predictions for the HSI22 and DB2 models appear as vertical straight lines in the $n_s$-$r$ plane. Similar to the GNI case, the curves for Nieh-Yan modified gravity lie on top of the standard Einstein gravity predictions with $f=0.1\,\Mpl$ and $n \in [0.01, 99]$ for HSI22, $n \in [50, 99]$ for DB2.\footnote{Recall that $M  = f\sqrt{n}$, so that these parameter choices restrict $M < M_{\rm Pl}$.} The HSI22 model yields $n_s \approx 0.965$, satisfying the Planck constraints ($n_s \simeq 0.9649 \pm 0.0042$) \cite{Planck:2018jri}. In contrast, the DB2 model predicts $n_s \approx 0.975$, which satisfies constraints from the Atacama Cosmology Telescope ($n_s \simeq 0.9743 \pm 0.0034$) \cite{calabrese2025atacamacosmologytelescopedr6}. We also see that GNI0.1 satisfies both Planck constraints \cite{Planck:2018jri} and constraints from the Atacama Cosmology Telescope \cite{calabrese2025atacamacosmologytelescopedr6}. 

In general, we find that models with mildly super-Planckian decay constants, $f \gtrsim M_{\rm Pl}$, that satisfy observational constraints in standard General Relativity, can be made to satisfy constraints with much smaller decay constants by coupling to the Nieh-Yan term. Since Cosine Natural inflation ($p = 1$ in eq \eqref{general natural inflation}) remains in tension with current observational constraints due to the amplitude of its tensor fluctuations (recall BICEP-Keck data \cite{BICEP:2021xfz} constrains $r < 0.036$ while Planck is best fit by $n_s \simeq 0.96-0.97$), we do not consider it in this section.

\begin{figure}[t]
  \centering
  \includegraphics[width=\linewidth]{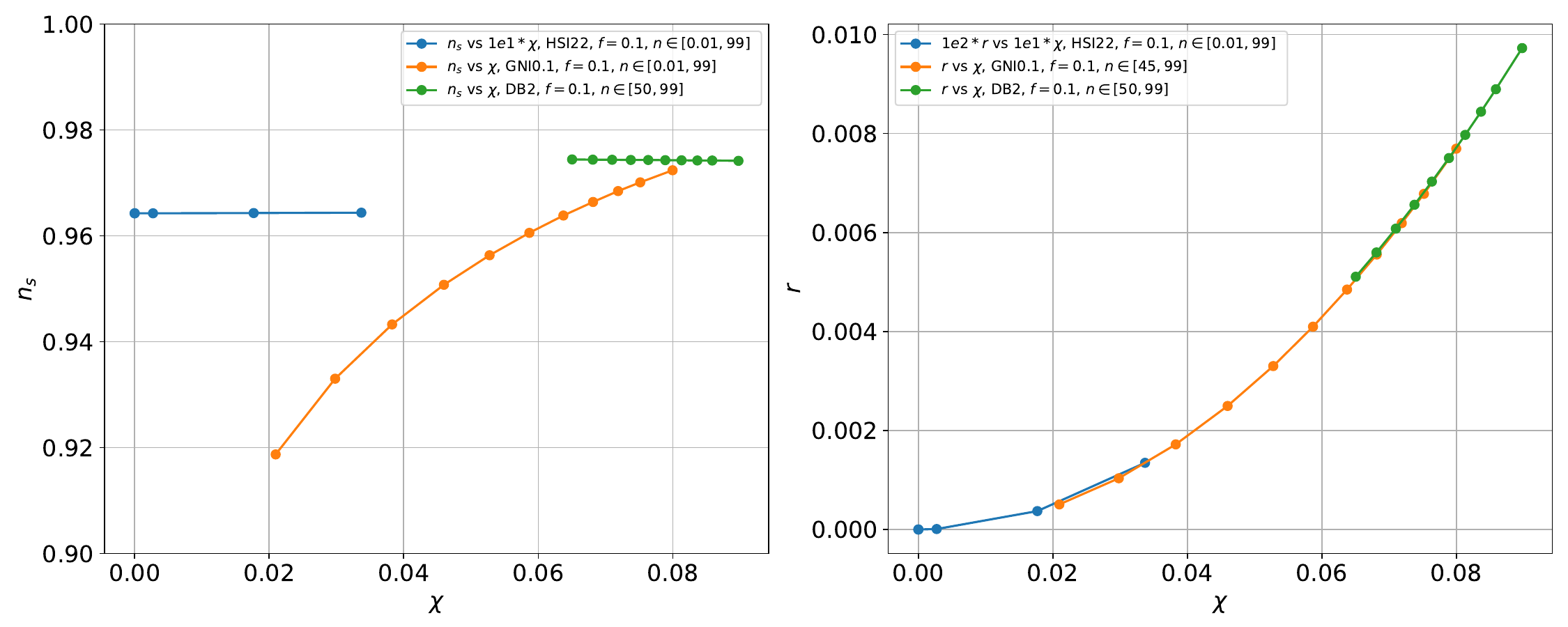}
  \caption{$n_s$ and $r$ vs $\chi$ for HSI$22$, GNI$0.1$ and DB$2$ model, $\Mpl = 1$}
  \label{fig chi vs ns and r sqh cni nieh yan}
\end{figure}

Furthermore, from figure \ref{fig chi vs ns and r sqh cni nieh yan}, we find that, based on the constraints on $n_s$ and $r$ \cite{Planck:2018jri,   calabrese2025atacamacosmologytelescopedr6,  BICEP:2021xfz}, the chirality parameter is predicted to be $\chi \leq 0.0034$,\footnote{Notice that in figure \ref{fig chi vs ns and r sqh cni nieh yan}, we plot $n_s$ vs $10\times\chi$ and $100\times r$ vs $10\times\chi$ for HSI22 model. } $0.059 \leq\chi \leq 0.08$, and $\chi \leq 0.095$  for the HSI22, GNI0.1, and DB2 models, respectively, within Nieh-Yan modified gravity. While this provides an observation signature of our model, detecting chirality at this (or any) level is likely to be challenging in upcoming cosmic microwave background (CMB) experiments \cite{Gluscevic:2010vv, Caldwell:2017chz}. However, we anticipate that parity violation is also likely to be manifest in higher-order correlators, such as the graviton-mediated trispectrum \cite{Seery:2008ax, Fujita:2023inz}. 

%
\section{Conclusions}\label{section conclusions}
%

In this paper, we have studied inflation in theories with torsion. We worked in the Einstein-Cartan-Palatini (first-order) formulation of General Relativity, and allowed for torsion by treating the spin connection as an independent degree of freedom.  We considered two modified gravity theories: a pseudoscalar coupled to the Pontryagin and Nieh-Yan densities. In the presence of these terms, the rolling pseudoscalar can source a non-trivial torsion background during inflation. For dynamical Chern-Simons gravity, a very large non-minimal coupling between the axion and the Pontryagin density is required to support a significant torsion field. For Nieh-Yan modified gravity, we found that the effect of the torsion field was to effectively increase the value of the axion decay constant $f\to f\sqrt{1+6M^4/\Mpl^2 f^2}$, allowing for viable inflation models at parametrically small values of the axion decay constant. A sub-Planckian decay constant is mapped to an effectively super-Planckian value due a hierarchy of scales between the scale of new physics associated with the Nieh-Yan term, $M$, which we anticipate to be near the Planck scale, and the axion decay constant which should be $f \ll \Mpl$.

We studied linear tensor perturbations about these inflationary background solutions, and found that, for Chern-Simons modified gravity, ghost and gradient instabilities appear on super-horizon scales and render the theory inviable for inflation.   In Nieh-Yan modified gravity, the parity-violating torsion field induces a difference between the two helicity modes of gravitational waves, leading to a chiral primordial tensor power spectrum, while having only a mild effect on its amplitude. 

In the scalar sector, for Nieh-Yan modified gravity, we found that the remapping of the axion decay constant holds even after accounting for all gravitational constraints. Due to the fact that the scalar power spectrum is proportional to $\delta\vartheta/\dot{\vartheta}$, the remapping effectively cancels out, and our model predicts approximately the same scalar spectrum as the analogous inflation model in general relativity, however, with a decay constant remapped to $f \to f\sqrt{1+6M^4/\Mpl^2 f^2}$. This fact, in combination with the fact that the broad features of the tensor spectrum are unchanged, means that our model does not dramatically alter the landscape of existing models for pseudoscalar inflation. However, our model does appear to allow existing models, that might otherwise require uncomfortably large axion decay constants, to work with much smaller axion decay constants. Thus, allowing for torsion and coupling to the Nieh-Yan density may potentially allow one to sidestep the issues associated with the compatibility of large axion decay constants with either the weak gravity conjecture or with string theory.

Our study has been purely phenomenological. We have assumed that the required Nieh-Yan density can be generated with a new mass scale at or near the Planck scale; detailed constructions that show this is possible are beyond the scope of this work. Furthermore, we have restricted ourselves to linear perturbations about the cosmological background. Beyond linear order, the non-Gaussian signatures in the scalar and tensor sectors may provide additional phenomenology that further distinguishes (or constrains) our scenario. For example, we anticipate that our model will generate a parity-violating trispectrum through graviton exchange (see, for example, \cite{Fujita:2021flu}). We leave studies of scalar, tensor and mixed non-Gaussianities to future work. Beyond cosmological fluctuations, in our model, the background torsion field provides a classical source for the Pontryagin density. This may provide a natural mechanism for leptogenesis via the gravitational anomaly in the standard model lepton current \cite{Alexander:2004us, Maleknejad:2014wsa, Maleknejad:2016dci, Caldwell:2017chz, Adshead:2017znw}. Finally, we considered only non-dynamical torsion in this work. Another interesting avenue would be to study the dynamics of the system with dynamical torsion (see, for example, \cite{Benisty:2021sul}).  We leave exploration of these directions to future work.

\section*{Acknowledgments}

We thank Stephon Alexander, Rob Leigh, and Nicolas Yunes for useful conversations. The work of P.A. and I.D. is supported in part by the United States Department of Energy through grant DE-SC0015655. SB is supported in part by the Higgs Fellowship and by the STFC Consolidated Grant ``Particle Physics at the Higgs Centre''.

\appendix
%
\section{Ansatz for the torsion tensor}\label{Appendix reduction of the general torsion tensor to the effective one}
%

We introduce the most general ansatz for the torsion 2-form\footnote{Note that a similar form for the torsion tensor was considered in the context of Poincare Gauge Gravity in reference \cite{Lu:2016bcx}}
\bea\label{appendix general ansatz torsion}
T^0 &=& 0,
\\
T^i &= & h(t)e^0\wedge e^i + h^i_{j}(t)e^0 \wedge e^j - \phi(t)\epsilon^i_{jk} e^j \wedge e^k+f^i{}_{l}\epsilon^l_{jk} e^j \wedge e^k,
\eea
where $h^{ij}$ and $f^{il}$ are traceless.  The torsion-full parts of the spin-connection, $\tilde\omega^{AB}$, are
\bea
\tilde\omega^{0i} &=& h(t)e^i - \frac{h^{ij}(t)}{2}e^j\,,\\
\tilde\omega^{ij} &=&  \frac{h^{ij}(t)}{2}e^0 +\phi(t)\epsilon^{ij}{}_{k}e^k - f^i_l \epsilon^{lj}{}_{k}e^k\,.
\eea
Since the spin connections should be antisymmetric, we  have
\begin{align}
h^{ab} &= -h^{ba},
\\
f^{ac} \epsilon^{cbd} &= -f^{bc} \epsilon^{cad},
\end{align}
which implies that the diagonal elements of $h^{ab}$ are $0$, the off-diagonal elements of $f^{ab}$ are $0$ and $f^{11} = f^{22} = f^{33}$. This, combined with the traceless condition, requires $f^{11} = f^{22} = f^{33} = 0$. Hence, $f^{ab} = 0$. The torsion constraint, eq.\ \ref{torsion constraint CS and NY equation}, for both the dynamical Chern-Simons and Nieh-Yan cases, gives
\begin{align}
h^{ab} = h^{ba}
\end{align}
Hence $h^{ab} = 0$. Thus, the torsion 2-form ansatz \ref{appendix general ansatz torsion} reduces to
\bea \label{ansatz torsion}
T^0 &=& 0,
\\
T^i &= & h(t)e^0\wedge e^i - \phi(t)\epsilon^i_{jk} e^j \wedge e^k.
\eea
In the tensor sector, the 2-form torsion ansatz, eq.\ \ref{appendix general ansatz torsion}, gets tensor perturbations
\bea\label{appendix general ansatz torsion tensor perturbation}
h^{ij} &=& \tilde{h}^{ij} + h_t^{ij},
\\
f^{il} &=& \tilde{f}^{il} + f_t^{il},
\eea
where $\tilde{h}^{ij}$ and $\tilde{f}^{il}$ are $0$ and $h_t^{ij}$ and $f_t^{il}$ are traceless. The antisymmetric property of the spin connections and the torsion constraint, eq.\ \eqref{torsion constraint CS and NY equation}, for both Dynamical Chern-Simons and Nieh-Yan, gives us $th^{ij} = 0 = tf^{il}$.

In the scalar sector, the 2-form torsion ansatz, eq.\ \eqref{appendix general ansatz torsion}, gets scalar perturbations
\bea\label{appendix general ansatz torsion tensor perturbation}
h^{ij} &=& \tilde{h}^{ij} + \delta^{ij}h_{s} + \partial^i \partial^j h_{ss}
\\
f^{il} &=& \tilde{f}^{il} + \delta^{il}f_{s} + \partial^i \partial^l f_{ss}
\eea
where $\tilde{h}^{ij}$ and $\tilde{f}^{il}$ are $0$. Again, the antisymmetric property of spin connection will require $h_{s} = 0 = f_{s}$, $\partial_l{f_{ss}} = \text{constant}$ and $\partial_j{h_{ss}} = \text{constant}$. The torsion constraint, eq.\ \eqref{torsion constraint CS and NY equation} for Nieh Yan, gives $f_{ss} = 0$ while the Einstein equation, eq.\ \eqref{einstein equation CS and NY} for Nieh Yan, gives $h_{ss} = 0$.

%
\section{Tensor power spectrum}\label{Appendix Expression for Tensor power spectrum}
%

In conformal time ($\eta$), the equation of motion from the tensor perturbed action for Nieh-yan modified gravity gives:
\begin{equation}
\frac{\partial^2 u_{\mathbf{k}}^{\pm}}{\partial x^2} + \(1 \mp \frac{4\phi}{Hx} - \frac{2 + 3\epsilon_H}{x^2}\)u_{\mathbf{k}}^{\pm} = 0
\end{equation}
This is the Coulomb wave equation
\begin{equation}
\frac{d^2w}{d\rho^2}+\(1 - \frac{2\beta}{\rho} - \frac{l(l+1)}{\rho^2} \)w = 0
\end{equation}
whose general solutions are Coulomb wave functions or the Coulomb-Hankel function, $H^{\pm}_{l}(\beta,\rho)$. As $|\rho| \rightarrow \infty$, $H^{\pm}_{l}(\beta,\rho) \simeq \exp [\pm\theta_{l}(\beta,\rho)]$, where $\theta_{l}(\beta,\rho) = \rho - \beta \ln (2\rho) - \frac{1}{2}l\pi + \sigma_{l}(\beta)$ and $\sigma_{l}(\beta) = \text{ph} \Gamma(l+1+i\beta)$. $H_{l}^{+}(\beta,\rho)$ and $H_{l}^{-}(\beta,\rho)$ are also complex conjugates.

In our case, $\rho = x$, $l = 1 + \epsilon_H$ and $\beta = \pm \frac{2\phi}{H}$. Imposing Bunch Davies vacuum conditions as $x \rightarrow -\infty$, picks out the solution
\begin{align}
H_{l}^{-}(\beta,\rho) = G_{l}(\beta,\rho) - iF_{l}(\beta,\rho)
\end{align}
where $G_{l}(\beta,\rho)$ and $F_{l}(\beta,\rho)$ are Irregular and Regular Coulomb wave functions respectively. Therefore, the tensor mode functions are
\begin{equation}
u^{\pm}_{\mathbf{k}} = C_{\pm}H_{1+ \epsilon_H}^{-}\(\pm \frac{2\phi}{H},k\eta\) = C_{\pm}\[G_{1+ \epsilon_H}\(\pm \frac{2\phi}{H},k\eta\)-  iF_{1+ \epsilon_H}\(\pm \frac{2\phi}{H},k\eta\)\]
\end{equation}
and
\begin{equation}
u^{\pm}_{\mathbf{k}}\hspace{1pt}^{*} = C_{\pm}^{*}H_{1+ \epsilon_H}^{+}\(\pm \frac{2\phi}{H},k\eta\) = C_{\pm}^{*}\[G_{1+ \epsilon_H}\(\pm \frac{2\phi}{H},k\eta\) +  iF_{1+ \epsilon_H}\(\pm \frac{2\phi}{H},k\eta\)\]
\end{equation}
The modes are canonically quantized, provided the Wronskian ($W$) condition is satisfied
\begin{equation}
    W[u_{\mathbf{k}}^{\pm},u_{\mathbf{k}}^{\pm}\hspace{1pt}^{*}] = i
\end{equation}
Substituting the expressions for the mode functions, we obtain
\begin{equation}
W[u_{\mathbf{k}}^{\pm},u_{\mathbf{k}}^{\pm}\hspace{1pt}^{*}] = 2ik|C_{\pm}|^{2},
\end{equation}
giving
\begin{equation}
   C_{\pm} = \frac{1}{\sqrt{2k}}
\end{equation}
Hence,
\begin{equation}
u^{\pm}_{\mathbf{k}} = \frac{1}{\sqrt{2k}} H_{1+ \epsilon_H}^{-}\(\pm \frac{2\phi}{H},k\eta\)
\end{equation}
In the limit $\rho \rightarrow 0$:
\begin{equation}
F_{l}(\beta,\rho) \sim C_{l}(\beta)\rho^{l+1} \hspace{5pt},\hspace{15pt} G_{l}(\beta,\rho) \sim \frac{\rho^{-l}}{(2l+1)C_{l}(\beta)}
\end{equation}
where $C_{l}(\beta) = \frac{2^{l}e^{-\pi \beta/2}|\Gamma(l+1+i\beta)|}{(2l+1)!}$. Therefore, in our case, as $x \rightarrow 0$,
\begin{equation}
|u^{\pm}_{\mathbf{k}}|^2 = \frac{1}{2k}\[\frac{(k\eta)^{-2}}{(3C_{1 + \epsilon_{H}}(\pm \frac{2\phi}{H}))^2}\]
\end{equation}

%
\section{Details of the scalar perturbations}\label{Appendix Equation of motion for the scalar perturbation}
%
We set $\Mpl = 1$ here, but it has been restored in section \ref{subsection scalar perturbation in Nieh Yan} and \ref{subsection scalar power spectrum}. In spatially-flat gauge, the metric is given by
\begin{equation}
[g_{\mu\nu}] = a^{2}(\eta) [\eta_{\mu\nu} + h_{\mu\nu}] =
  a^{2}(\eta)\begin{bmatrix}
   1+2A &
   -\partial_{i}B \\
   -\partial_{i}B &
   -\delta_{ij} 
   \end{bmatrix}
\end{equation}
with inverse
\begin{equation}
[g^{\mu\nu}] = a^{-2}(\eta) [\eta^{\mu\nu} - h^{\mu\nu}] =
  a^{-2}(\eta)\begin{bmatrix}
   1-2A &
   -\partial_{i}B \\
   -\partial_{i}B &
   -\delta_{ij} 
   \end{bmatrix}
\end{equation}
The vielbeins/tetrads ($e^{A}\hspace{0.5pt}_{\mu}$) are taken to be
\begin{align}
e^{0}\hspace{0.5pt}_{0} =  a[1+A], \quad 
e^{0}\hspace{0.5pt}_{i} =  a\partial_{i}\beta, \quad 
e^{a}\hspace{0.5pt}_{0} =  a\delta^{ai}\partial_{i}\zeta, \quad
e^{a}\hspace{0.5pt}_{i} =  a[\delta_{ia} + \epsilon_{aik}\partial_{k}s].
\end{align}
and $e_{A}\hspace{0.5pt}^{\mu}$
\begin{align}
e_{0}\hspace{0.5pt}^{0} =  a^{-1}[1-A], \quad 
e_{0}\hspace{0.5pt}^{i} =  -a^{-1}\partial_{i}\zeta, \quad 
e_{a}\hspace{0.5pt}^{0} =  -a^{-1}\delta^{i}_{a}\partial_{i}\beta, \quad 
e_{a}\hspace{0.5pt}^{i} =  a^{-1}[\delta_{ia} - \epsilon_{aik}\partial^{k}s]
\end{align}
 and finally $e^{A\mu}$
\begin{align}
e^{00} =  a^{-1}[1-A], \quad 
e^{0i} =  -a^{-1}\partial_{i}\zeta, \quad 
e^{b0} = & a^{-1}\delta^{i}_{b}\partial_{i}\beta, \quad 
e^{bi} = & -a^{-1}[\delta_{ib} - \epsilon_{bik}\partial^{k}s]
\end{align}
The Christoffel symbols are given by
\begin{align}\nonumber
\Gamma^{0}_{00} = &  \mathcal{H} + A', \quad 
\Gamma^{0}_{0i} =  \partial_{i}A + \mathcal{H}\partial_{i}B, \quad 
\Gamma^{i}_{00} =  \partial_{i}B' + \mathcal{H}\partial_{i}B + \partial_{i}A
\\\nonumber
\Gamma^{i}_{j0} = & \mathcal{H}\delta_{ij}, \quad 
\Gamma^{0}_{ij} =  \mathcal{H}\delta_{ij} - 2\mathcal{H}A\delta_{ij} - \partial_{i}\partial_{j}B, \quad 
\Gamma^{i}_{jk} =  -\mathcal{H}\delta_{jk}\partial_{i}B
\end{align}
where a prime, $\;'\;$, denotes derivative with respect to conformal time ($\eta$) and $B = \zeta - \beta$.  

The ``torsion-less'' spin connections are:
\begin{align}\nonumber
\bar{\omega}_{0}\hspace{0.5pt}^{0b} = &
\delta^{bi}[ \partial_{i}\beta' - \partial_{i}A - \mathcal{H}\partial_{i}B ], \quad
\bar{\omega}_{i}\hspace{0.5pt}^{0b} =   -\mathcal{H}\delta^{b}_{i} + \mathcal{H}A\delta^{b}_{i} + \delta^{bj}\partial_{i}\partial_{j}\zeta + \mathcal{H}\epsilon_{ibj}\partial_{j}s\\
\bar{\omega}_{0}\hspace{0.5pt}^{ab} = &  \epsilon_{abk} \partial_{k}s', \quad\qquad
\bar{\omega}_{i}\hspace{0.5pt}^{ab} =  \mathcal{H}\delta^{a}_{i}\delta_{bj}\partial_{j}\beta - \mathcal{H}\delta^{b}_{i}\delta_{aj}\partial_{j}\beta + \epsilon_{abj}\partial_{i}\partial_{j}s
\end{align}
While the ``torsion-full'' spin connections are:
\begin{align}\nonumber
\tilde{\omega}_{0}\hspace{0.5pt}^{0a} = & he^{a}\hspace{0.5pt}_{0}, \quad \tilde{\omega}_{i}\hspace{0.5pt}^{0a} =  he^{a}\hspace{0.5pt}_{i}
\\\nonumber
\tilde{\omega}_{0}\hspace{0.5pt}^{ab} = & \phi \epsilon^{ab}\hspace{0.5pt}_{c}e^{c}\hspace{0.5pt}_{0}, \quad
\tilde{\omega}_{i}\hspace{0.5pt}^{ab} =  \phi \epsilon^{ab}\hspace{0.5pt}_{c}e^{c}\hspace{0.5pt}_{i}.
\end{align}
The equations of motion for the perturbations after varying the spin connection
\begin{align}\nonumber
2nfa^2 \vartheta' \partial_{i}s &= 2a^3\phi\partial_{i}s
\\\nonumber
2nfa^2 \delta \vartheta' &= 2a^3(\delta\phi + A\phi)
\\\nonumber
2nfa^2 \partial_{i}\delta \vartheta &= 2a^3 \phi \partial_{i}\beta
\\
0 &= 2a^3 \delta h
\end{align}
with solutions
\begin{align}\nonumber
\delta h = & 0,
\\\nonumber
\delta\phi = & \frac{nf\delta\vartheta'}{a} - \frac{nf\vartheta'}{a}A,
\\
\beta = & \frac{\delta\vartheta}{\vartheta'}.
\end{align}
Einstein's equation gives us the necessary relationship for $\nabla^2 B$ and $A$:
\begin{align}\nonumber
2\mathcal{H}\partial_{i}A - \frac{3(\mathcal{H}^2 + n^2f^2\vartheta'^2)}{\vartheta'}\partial_{i}\delta\vartheta = & - \frac{(2a^2 V - \vartheta'^2)}{2\vartheta'}\partial_{i}\delta\vartheta,
\\
-2[3A(\mathcal{H}^2 - n^2f^2\vartheta'^2) + 3n^2f^2\vartheta'\delta\vartheta' + \mathcal{H}\nabla^2 B] = & a^2\frac{dV}{d\vartheta}\delta\vartheta + \vartheta'(-A\vartheta' + \delta\vartheta').
\end{align}
These give us
\begin{equation}
A = (1 + 6n^2f^2)\frac{\vartheta'\delta\vartheta}{2\mathcal{H}},
\end{equation}
and 
\begin{equation}
\nabla^2 B = -\frac{1}{2\mathcal{H}}\Big[ (1 + 6n^2f^2)\vartheta'\delta\vartheta' + a^2\frac{dV}{d\vartheta}\delta\vartheta + 2Aa^2V \Big].
\end{equation}
The Klein Gordon equation gives (after substituting $\phi$, $\delta\phi$ and $\beta$)
\begin{align}\nonumber
a^{2}\frac{dV}{d\vartheta} + (1+6n^2f^2)(\vartheta'' + 2\mathcal{H}\vartheta') = 0
\\\nonumber
a^{2}\frac{d^2V}{d\vartheta^2}\delta\vartheta - (1+6n^2f^2)(-\delta\vartheta'' - 2\mathcal{H}\delta\vartheta' + \nabla^{2}\delta\vartheta + \vartheta'\nabla^{2}B + \vartheta'A')
\\
- 2A(1+6n^2f^2)(\vartheta'' + 2\mathcal{H}\vartheta')= 0
\end{align}
Substituting the unperturbed equation for the second term in the perturbed equation and rearranging, we get
\begin{equation}
\delta\vartheta'' + 2\mathcal{H}\delta\vartheta' - \nabla^{2}\delta\vartheta - \vartheta'\nabla^{2}B - \vartheta'A' + \frac{a^{2}}{(1 + 6n^2f^2)}\frac{d^2V}{d\vartheta^2}\delta\vartheta + \frac{2Aa^{2}}{1 + 6n^2f^2}\frac{dV}{d\vartheta} = 0.
\end{equation}
Substituting $A$ and $\nabla^2 B$ and switching to Fourier space, we arrive at
\begin{align}\nonumber
\delta \vartheta'' + 2\mathcal{H}\vartheta' + k^2 \delta\vartheta + \frac{a^2}{1 + 6n^2f^2} \frac{d^2V}{d\vartheta^2}\delta\vartheta
- \vartheta'(1+ 6n^2f^2)\[ \frac{\vartheta''\delta\vartheta}{\mathcal{H}} + \frac{\vartheta'\delta\vartheta'}{\mathcal{H}} - \frac{\vartheta'\delta\vartheta\mathcal{H}'}{\mathcal{H}^2} \]
\\
+ \frac{\vartheta'}{2\mathcal{H}}\[ (1 + 6n^2f^2)\vartheta'\delta\vartheta' + a^2\frac{dV}{d\vartheta}\delta\vartheta + (1+6n^2f^2)\frac{2\vartheta'\delta\vartheta}{\mathcal{H}}a^2V \] + a^2\frac{dV}{d\vartheta} \frac{\vartheta' \delta\vartheta}{\mathcal{H}} = 0.
\end{align}

\bibliographystyle{JHEP}
\bibliography{main}

\end{document}